\documentclass[aps,prd,reprint,superscriptaddress,showpacs,nofootinbib]{revtex4-1}

\usepackage{comment}
\usepackage{amsmath}
\usepackage{multirow}
\usepackage{amssymb}
\usepackage{graphicx}
\usepackage{color}
\usepackage{bm}
\usepackage{gensymb}
\usepackage{bigints}
\usepackage{placeins}
\usepackage[usenames,dvipsnames,svgnames,table]{xcolor}
\usepackage{pifont}
\usepackage{aas_macros}
\usepackage{orcidlink}
\usepackage{physics}
\usepackage{ragged2e}
\usepackage{bigints}
\usepackage{ctable}
\usepackage{subcaption}

\usepackage[labelfont=bf]{caption}

\hypersetup{
    colorlinks=true,
    linkcolor=Blue,
    filecolor=Blue,
    urlcolor=MidnightBlue,
    citecolor=Blue
}

\numberwithin{equation}{section}

\begin{document}

\title{Anisotropic hybrid stars: \\ Interplay of superconductivity and magnetic field leading to gravitational waves}

\author{Zenia Zuraiq\orcidlink{0009-0000-6980-6334}}
\email{zeniazuraiq@iisc.ac.in}
\affiliation{Department of Physics, Indian Institute of Science,
Bangalore 560012, India}

\author{Banibrata Mukhopadhyay\orcidlink{0000-0002-3020-9513}}
\email{bm@iisc.ac.in}
\affiliation{Department of Physics, Indian Institute of Science,
Bangalore 560012, India}

\begin{abstract}
\noindent

Neutron stars, at their cores, are highly dense and, thus, are expected to have a number of exotic processes. This includes a possible phase transition to deconfined quark matter at the core, leading to a hybrid star. The quark matter is expected to additionally be color superconducting.
The physics of superconductivity plays an important role in understanding the high density matter in the interiors of neutron/hybrid stars. At their high densities, additionally, both proton superconductivity and neutron superfluidity are expected. We study the effect of superconducting (quark/proton) matter, along with the internal magnetic field, leading to pressure anisotropy within hybrid stars. We aim to probe the effect of superconductivity, especially from color superconducting quarks, on hybrid star structure. We propose new phenomenological model anisotropy profiles within a one-dimensional framework. We model quark matter using the vector interaction enhanced Bag model, and hadron matter with the DD2 equation of state. A Maxwell construction joins both phases. We further investigate the possible observational signatures of these hybrid stars. These include mass enhancement and the possibility of continuous gravitational waves arising from the anisotropy induced deformation, giving us a way to further constrain our model and its physical parameters.

\end{abstract}

\maketitle

\section{Introduction }\label{Xsec1-1}
\label{sec:intro}
Neutron stars (NSs) are among the most theoretically rich objects in our Universe. The cores of NSs are the sites of extremely dense matter - several times denser than the densities inside atomic nuclei. In these highly dense regions, a number of exotic processes could arise. Among these are the generation of non-terrestrially stable particles such as hyperons, boson condensates (pions, kaons), etc. One such suggestion is the possible appearance of deconfined quark matter \cite{weber2005} at high densities  - leading to the formation of hybrid stars (HSs): neutron stars with deconfined quark cores. An even more exotic scenario \cite{bodmer,witten} is the emergence of absolutely stable strange matter within the NS core - this leads to the formation of strange quark stars (SQSs).

An important aspect in this discussion is the fact that these are \textit{low temperature} systems with Fermi temperature several orders of magnitude higher than the matter temperature $T$. Hence the cores of NSs are very well approximated as cold (with $T=0 \ K$) gases. Within a few seconds of their birth, NSs are expected to reach temperatures below 10 MeV. Under such conditions, any deconfined quarks may additionally exist in color superconducting (CSC) phases. Owing to the uncertainties in QCD physics at intermediate densities, the exact nature and pairing pattern of the CSC phases is still unknown (for a review of color superconductivity physics, see \cite{rajagopal_rev}).

The most symmetric among the CSC phases is the color flavor locked (CFL) phase, where quarks of all three colors and all three flavors participate in the pairing. It is readily achieved at ultra-high densities, where the quarks can be considered as completely massless. At HS core densities, the appearance of this phase is theorized based on its symmetric, low energy configuration. However, its appearance is contingent on the competition between the superconducting energy gap ($\Delta_{\text{CFL}} \approx 10 - 100$ MeV) and the strange quark mass ($M_s$). In particular, CFL is disfavored when $\Delta_{\text{CFL}} < M_s^2/4\mu$, as the stresses from the differences in Fermi momenta at the same chemical potential ($\mu$) for different quark flavors disfavor the Cooper pairing. In the absence of CFL pairing, less symmetric CSC phases such as 2SC, CSL, etc. could arise in the star.

It is important to note that \textit{proton} superconductivity (and neutron superfluidity) has long been theorized to exist inside NSs, irrespective of the presence of quark phases, with pairing gaps $\Delta_{\rm SC} \lesssim 1$ MeV. However, recent work has shown \cite{das_sc,sinha_sc_magnetars} that the fields at the cores of highly magnetized NSs (i.e., magnetars) exceed the critical magnetic field for this pairing, and thus, these objects may be devoid of proton superconductivity. Moreover, higher central densities leading to NSs more massive than $1.4M_\odot$ make $\Delta_{\rm SC}=0$. Notably, the critical fields required for destroying color superconductivity are much higher - this is due to the CSC gap parameter being 10--100 times larger than the proton superconducting gap. As a result, although massive magnetized NSs may have cores devoid of electromagnetic/proton superconductivity, they can still have CSC quark cores.

An interesting recent development in the field of NSs is the observations of objects in the ``lower mass gap''. This refers to objects which are increasingly being observed in the mass window between $2.5M_\odot$ and $5M_\odot$, with uncertainty of whether they are in fact NSs or light black holes. For instance, the secondary object observed in the gravitational wave event GW190814 \cite{massgap_GW190814} was inferred to have a mass within the range of $2.50 - 2.67M_\odot$. Additionally, one of the components of the eccentric binary PSR J0514-4002E has an inferred mass within the mass gap \cite{massgap_Sci}. Other gravitational wave observations including GW230529 \cite{GW230529} also suggest the possibility of objects in the mass gap. On the other hand, high mass pulsar observations, e.g., PSR J0952-0607 of mass $2.35 \pm 0.17M_\odot$ \cite{j0952_1,j0952_2} and NICER observations, e.g., PSR J0740+6620 of mass $2.08 \pm 0.07M_\odot$ \cite{j0740_1,j0740_2} show that the maximum mass ($M_{\rm {\max}}$) of a NS is $\gtrsim 2M_\odot$. These observations make the existence of high mass NS/HS, possibly in the lower mass gap, a viable one.

We have previously explored the interpretation of massive, magnetized NSs as possible candidates to explain the objects in the lower mass gap \cite{zuraiq}. It was found that through a combination of magnetic field and anisotropic effects, one could counteract effects such as ``hyperon softening'' and thus form sufficiently massive, stable NSs. The presence of deconfined quark phases is an intriguing one as it is often associated with a similar softening effect. We thus examine if magnetic field and anisotropy, along with the added microphysics of color superconductivity, can work together to overcome this effect. In other words - does the presence of a deconfinement phase tranistion (PT) rule out HSs as mass gap candidates? Additionally, does the anisotropy studied here and the subsequent deformation of the HS lead to observable signatures, e.g., continuous gravitational waves (CGWs)?

This work addresses the above questions by modeling the structure of anisotropic, superconducting, magnetized HSs. In Sec. \ref{sec:Formalism}, we summarize our formalism for the construction of the HSs. In Sec. \ref{sec:aniso}, we propose ways to tie the stellar anisotropy to microphysics of magnetic, superconducting HSs. In Sec. \ref{sec:varySC}, we further examine the extent/variation of various superconducting phases within the HS. We look for possible observable signatures in Sec. \ref{sec:obs}, including effects on $M_{\mathrm{max}}$ supported by the HS. We also explore the possibility of anisotropy induced deformation leading to CGWs. We end with our conclusions in Sec. \ref{sec:conclusion}.

\section{Formalism}\label{Xsec2-2} \label{sec:Formalism}

Following previous work \cite{deb1,deb2,zuraiq}, we model anisotropic compact stars in approximate spherical symmetry by modifying the Tolman-Oppenheimer-Volkoff (TOV) equations to incorporate the anisotropic and/or magnetic field effects. The stellar structure is then described by\footnote{We consider $G = c = 1$ units. },
\begin{equation} \label{tov1}
    \frac{dm}{dr} = 4 \pi r^2 \left(\epsilon + \frac{B^2}{8\pi}\right), \
\end{equation}
\begin{equation} \label{tov2}
    \frac{dp_r}{dr} =
    \begin{cases}
    \frac{- (\epsilon + p_r)\frac{\left(4\pi r^3\left(p_r - \frac{B^2}{8\pi}\right) + m \right)}{r(r-2m)} + \frac{2}{r}\sigma_{\rm aniso}}{\left(1 - \frac{d}{d\rho}\left(\frac{B^2}{8\pi}\right)\left(\frac{d\rho}{dp_r}\right)\right)} & \text{(for RO),} \\
    \frac{- (\epsilon + p_r + \frac{B^2}{4\pi})\frac{\left(4\pi r^3\left(p_r + \frac{B^2}{8\pi}\right) + m \right)}{r(r-2m)} + \frac{2}{r}\sigma_{\rm aniso}}{\left(1 + \frac{d}{d\rho}\left(\frac{B^2}{8\pi}\right)\left(\frac{d\rho}{dp_r}\right)\right)} & \text{(for TO)}.
    \end{cases}
\end{equation}

Here, $\sigma_{\rm aniso} = p_t -p_r$ is the pressure anisotropy at every point in the star, with $p_t$ being the pressure in the transverse direction. We parameterize this based on anisotropy profiles discussed in Sec. \ref{sec:aniso}. As explored in previous work \cite{deb1,deb2,zuraiq,zz_univ}, the contributions of the $B^2/8\pi$ terms to the pressures and densities captures the combined effect of magnetic pressure and magnetic tension in the radial direction. The anisotropy term $\sigma_{\rm aniso}$ allows us to further introduce transverse variations in pressure in the system within a one-dimensional (1D) framework.
These equations are further supplemented by the profiles for the magnetic field within the star and the equation(s) of state (EoS). We use a pressure-dependent parameterization for the field instead of the usual density dependence considered in previous work \cite{Bprof}. This allows us to study HSs where PTs can lead to large discontinuities in the density within the star, while keeping pressure continuous.

In the model, we capture different magnetic field geometries by considering both radially oriented (RO) and transversely oriented (TO) fields. These are the 1D analogs for poloidal and toroidal magnetic fields respectively. In the present work, we restrict our study to TO fields. It is well known that toroidal fields lead to a lesser degree of deformation, leading to the star maintaining approximate spherical symmetry \cite{pili}. This is in line with our 1D modeling. Further, certain stability studies indicate that the toroidal field could be stronger than the poloidal field, even up to 30 times larger \cite{braithwaite}. This justifies our study of TO stars, as we consider the structural properties of the star to be influenced by the dominant toroidal component. We do not explicitly consider the effects of Landau quantization, as these effects are expected to be dominant only for central magnetic fields $\gtrsim 10^{18} \ G$, beyond the fields considered in the current work \cite{mcfl,LandauEOS}. Magnetization is similarly practically ineffective at present, as its effect is expected to be an order of magnitude lower than the magnetic tension/pressure \cite{LandauEOS}.

The magnetic field profile is given by
\begin{equation}
\label{mag_bando}
    B(P) = B_s + B_0\left[1 - \exp\left\{-\eta{\left(\frac{p_r}{P_0}\right)^\gamma}\right\}\right],
\end{equation}
where, $B_s$, $B_0$, $\eta$, $P_0$ and $\gamma$ are model parameters. Here, $B_s$ is the surface field, $B_0$ controls the maximum (central) field, and $\eta$ and $\gamma$ control how the field decays from center to surface.

We consider $B_s = B_0/1000$, in line with the highest variation of magnetic field expected in realistic NSs \cite{zuraiq,das_sc}. As shown in previous studies using two-dimensional (2D) Einstein-Maxwell codes such as XNS (e.g. \cite{das_sc}), the toroidal component of the magnetic field could vary by two orders of magnitude from center to surface, while the poloidal/dipolar component only varies by an order of magnitude. Thus, by combining an internal toroidal field with a surface dipole field, the overall variation could easily be 3 orders as indicated in this work. For instance, consider a dominant toroidal field at the (near-)center of the star, 100 times the central dipole field. At the surface, the toroidal component drops to zero, and we only see the dominant dipole component, lower by an order of magnitude. Hence, there could be an overall variation of atmost three orders of magnitude.

In the current work, we fix $\eta = 0.01, \gamma = 1$ and $P_0= 2.5$ MeV/fm$^3$. The $P_0$ value corresponds to approximately the pressure at nuclear saturation density, $\rho_0= 0.153$ fm$^{-3}$, as predicted in various EoS models. The dependence of the structural properties of compact stars on $\eta$ and $\gamma$ has been studied elsewhere \cite{zuraiq,deb1,deb2}.

\subsection{Equation of state}\label{Xsec3-2.1}
An important ingredient in any NS/HS modeling is the EoS. Due to uncertainties in the physics of high density matter, the exact EoS is still unknown, with many models proposed for the same - both for nuclear matter and quark matter.

We use the DD2 EoS \cite{dd2} to describe the hadronic part of the EoS ($npe\mu$), constructed using density dependent relativistic mean field theory. The predictions of this EoS for the properties of symmetric nuclear matter (SNM) at nuclear saturation density ($n_0$) is given in Table \ref{tab:DDME2}.

\begin{table}[!htbp]
\begin{tabular}{r|l}
\hline
Property        &      \\ \hline
$n_0~({\rm fm}^{-3})$  & 0.152   \\ 
$E/N~({\rm MeV})$      & -16.14  \\ 
$J~({\rm MeV})$        & 32.3    \\ 
$L_0~({\rm MeV})$      & 51.25    \\ 
$J''~({\rm MeV})$      & -87.19   \\ 
$K~({\rm MeV})$        & 250.89   \\
$m_*/m$                & 0.572     \\ \hline
\end{tabular}

\caption{Properties of SNM at $n_0$ for the EoS used in this work. Properties tabulated here are - energy per nucleon ($E/N$), symmetry energy ($J$), slope of the symmetry energy ($L_0$), the curvature of the symmetry energy ($J''$), the incompressibility ($K$) and the effective mass ($m_*/m$).}

\label{tab:DDME2}
\end{table}

The quark EoS is described using the vector-interaction enhanced Bag (vBag) model \cite{vbag1,vbag2}. Here, the MIT bag model \cite{mit}, where quarks are considered to be ``free'' within a bag (characterized by the Bag constant $B_q$), is modified to include repulsive vector interactions (in terms of the parameter $K_v$). It also includes the effects of dynamical chiral symmetry breaking (D$\chi$SB). The inclusion of repulsive interactions is important to enable HSs to support masses of at least $2M_\odot$, as constrained by observations. The equation set corresponding to the vBag EoS is given as,
\begin{equation}
    \begin{aligned}
 &\mu_f = \mu^*_f + K_vn_{\rm FG}(\mu_f^*), \\&
 n_f(\mu_f) =  n_{{\rm FG}, f}(\mu^*_f), \\&
 P_f(\mu_f) =  P_{{\rm FG}, f}(\mu^*_f) + \frac{K_v}{2}n_{{\rm FG}, f}^2(\mu^*_f) - B_{q\chi,f}, \\&
  \epsilon_f(\mu_f) =  \epsilon_{{\rm FG}, f}(\mu^*_f) + \frac{K_v}{2}n_{{\rm FG}, f}^2(\mu^*_f) + B_{q\chi,f}, \\&
  P_{\rm Q} = \sum_f P_f + B_{dc}, \\&
\epsilon_{\rm Q} = \sum_f \epsilon_f - B_{dc},
   \end{aligned}
   \label{vbag_eq}
\end{equation}
where $\mu_f$ and $\mu_f^*$  are the physical and effective chemical potentials respectively. $n, P$ and $\epsilon$ have their usual meanings of number density, pressure and energy density respectively with the suffix $\rm FG$ denoting the free Fermi gas solution and $f$ can be $u$, $d$ or $s$ depending on the quark flavor under consideration. $B_{\chi,f}$ refers to the single flavor (chiral) bag constant, and $B_{dc}$ denotes the deconfinement bag constant. The effective bag constant of the theory is $B_{\rm eff} = \sum_{f=u,d,s} B_{\chi,f} - B_{dc}$.

Thus, we solve the equation set \eqref{vbag_eq} in the quark sector by additionally imposing charge conservation and chemical equilibrium conditions for the quark matter at every point, given below as,
\begin{equation}
    \begin{aligned}
 &\mu_u + \mu_e = \mu_d = \mu_s, \\&
 (2/3)n_u - (1/3)n_d - (1/3)n_s - n_e = 0. \\
   \end{aligned}
   \label{charge_neutrality}
\end{equation}

We use a Maxwell construction in order to join the hadronic and quark EoS to construct the HS EoS. The PT from hadronic matter to deconfined quark matter is characterized by a sharp, first-order phase boundary between the two phases. Note that, in general, even moderate values of surface tension at the interface disfavor the formation of a mixed phase (Gibbs construction) between the hadronic and deconfined quark phases \cite{CFL_ST}. At the point of PT, there is a discontinuity in the energy density $\Delta \epsilon$ while the pressure $P$ and chemical potential $\mu$ remain constant/equal in both phases and across the transition.

It is important to note here that 2D axisymmetric codes (e.g. XNS; \cite{xns}) use pseudo-enthalpy $H = \int_0^P dP'/(\epsilon(P')+ P')$ as the parameter for calculating their numerical solutions \cite{pseudoenthalpy}\footnote{The pseudo-enthalpy formalism as used in XNS is discussed at \url{https://www.arcetri.inaf.it/science/ahead/XNS/XNS4/eos.html}.}. In the presence of constant pressure phases associated with sharp first-order PTs, this approach may have issues, and hence we can focus on 1D approximations such as those used in the present work.

To include the effects of color superconductivity in the EoS, we note that the thermodynamic potential $\Omega_{\rm CFL}$ picks up an extra term due to diquark condensation such that $\Omega_{\rm CFL} = \Omega_{\rm unpaired} + \Omega_{\rm CSC}$ where $\Omega_{\rm CSC} = -(3/\pi^2) \Delta_{\rm CFL}^2 \mu^2$ is the CSC contribution, and $\Omega_{\rm unpaired}$ is the non-CSC contribution to the potential, coming from the vBag model in the present discussion \cite{lugones, reddy:CSC}. Here, the extra term particularly corresponds to CFL color superconductivity; other CSC phases have a similar modification, i.e., $\simeq \Delta_{\mathrm{SC}}^2\mu^2$, with a different constant factor. For instance, for the two-flavor pairing pattern of 2SC model, $\Omega_{\rm CSC}= -(1/\pi^2) \Delta_{\rm 2SC}^2 \mu^2$ (see \cite{schmitt_rev} for a detailed discussion on condensation/pairing energies of CSC matter). In the case of CFL color superconductivity, $\mu = \mu_B/3$ where $\mu_B = \mu_u + \mu_d + \mu_s$, i.e., the contributions are included from each quark flavor. From this thermodynamic potential, i.e. $\Omega_{\rm CFL}$, we derive all other thermodynamic quantities of the star. In the CFL case, this leads to an additional term $P_{\rm CSC} = -\Omega_{\rm CSC} = (3/\pi^2) \Delta_{\rm CFL}^2 \mu^2$, representing the CFL pairing energy, being added to the pressure of the vBag model. A corresponding $(2/\pi^2) \Delta_{\rm CFL}^2 \mu$ is added to the number density, derived from $n_{\rm CSC} = -\partial \Omega_{\rm CSC}/\partial\mu_B$. The additional contribution to the energy density is calculated as $\epsilon_{\rm CSC} = \mu_B n_{\rm CSC} - P_{\rm CSC} = (3/\pi^2) \Delta_{\rm CFL}^2 \mu^2$ \cite{lugones}. In the present work, as a first approximation, we consider the CSC phase to follow CFL type behavior and condensation energies. Other CSC phases may be treated in a similar manner by scaling the constant factor based on the number of quark flavors taking part in the pairing, leading to a change within the same order of magnitude.

\begin{figure}[!htpb]
	\centering
	\includegraphics[scale=0.55
    ]{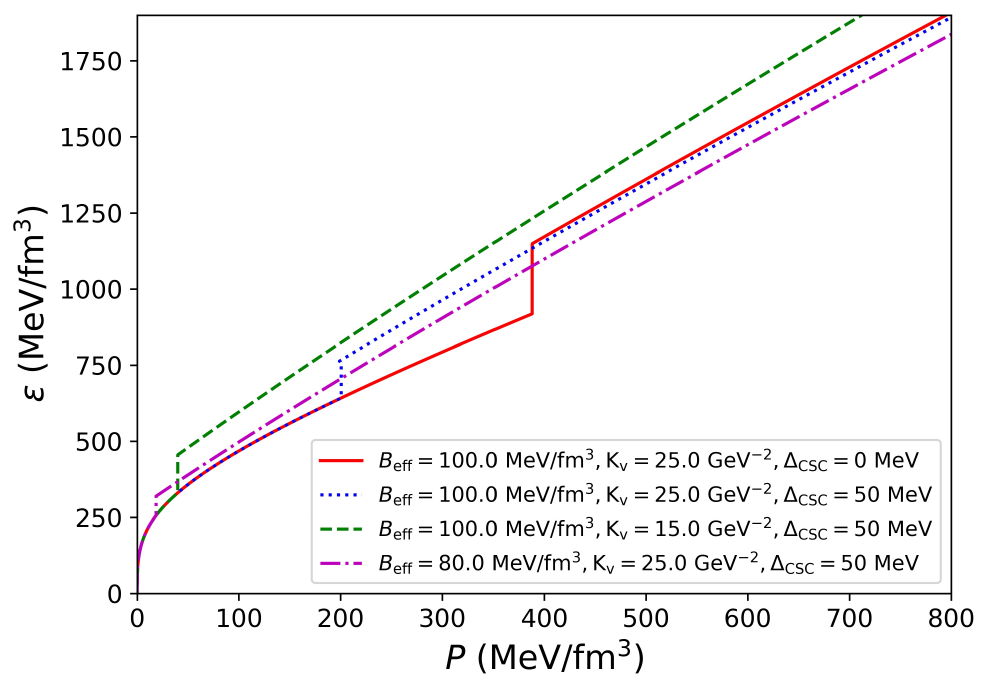}
	\caption{\justifying{Some representative hybrid EoS considered in this work, with corresponding $B_{\rm eff}, \ K_v$ and $\Delta_{\rm CSC}$.}}
    \label{fig:EOS}
\end{figure}

Some of the hybrid EoS used in this work are shown in Fig. \ref{fig:EOS}. Color superconductivity introduces an extra pairing energy term to the quark EoS, in both the pressure and energy density terms, and hence the $P$ at a given $\mu$ increases. This leads to the point of PT from hadron matter to quark matter being shifted to lower pressures, as shown in the figure. The effects of $B_{\rm eff}$ and $K_v$ on the point of PT are also illustrated in the figure. Notably, increasing either $B_{\rm eff}$ or $K_v$ leads to an increase in the point of PT. The shift of PT due to the interplay of these various parameters, and the resulting effect on the stellar structure and superconductivity, will be explored in further sections.

\section{Modeling anisotropy in superconducting, magnetized hybrid stars }\label{Xsec4-3}\label{sec:aniso}
We have previously used a modified phenomenological prescription to describe pressure anisotropy within NSs \cite{zuraiq,deb1,BL}. Similar modeling has been used in previous studies of HSs as well \cite{aniso_HS_BL,aniso_HS_QL}. In the present work, we propose two new phenomenological profiles, directly tied to the physical variables in the star. These profiles capture the extremes of how pressure anisotropy could be sourced by CSC matter within HSs. They can serve as alternates (which may be better for HSs) to the already proposed phenomenological anisotropy profiles present currently (e.g. \cite{BL, QLaniso}), with profiles that are directly tied to the physical effects of magnetic fields and superconductivity, expected within HSs.

To start with, magnetic fields naturally lead to anisotropy in NSs/HSs. This leads to the differing contributions of magnetic pressure and magnetic tension to the pressures in the radial and transverse directions.
The presence of superconductivity - both color superconductivity and proton superconductivity - can further affect the magnetic stresses and anisotropic pressures in the star. It has long been known, for instance, that in the presence of type II proton superconductivity, magnetic stresses are further enhanced by a factor of $H_{c1}/B$ \cite{type2_stress,Hc1B_ellipticity}, where $H_{c1}$ is the first critical magnetic field, above which the magnetic flux is threaded through vortex tubes. This critical magnetic field is a function of the superconducting gap within the star. The presence of a (strong) magnetic field can also affect CSC states - e.g., leading to different symmetry patterns such as the MCFL phase \cite{mcfl}, which leads to an enhancement in the field through enhanced magnetization \cite{mcfl_magnetars}, dependent on the superconducting gap.
    Thus, the presence of superconductivity may lead to a further enhancement of magnetic stresses and anisotropy in the star. The strength of this modification is directly/indirectly dependent on the superconducting gap ($\Delta_{\rm SC}$), arising from either proton or color superconductivity. Further, as discussed previously, the pairing/condensation energy associated with superconductivity is $\sim \Delta_{\rm SC}^2\mu^2$ \cite{schmitt_rev} and this sets the energy scale for pressure/EoS modifications due to superconductivity.

Based on all these factors, we propose the following anisotropy profile in the star (hereinafter, Profile 1),
\begin{equation}
\sigma_{\rm aniso} = p_t - p_r = \pm  B^2\left(\frac{2m}{r}\right)\left(1 + \frac{\Delta_{\rm SC}^2\mu^2}{p_r}\right)\left(\frac{p_r}{p_c}\right),
\label{Xeqn6-3.1}
\end{equation}
where $p_t, p_r \text{ and } p_c$ are the transverse, radial and central pressures of the HS respectively, $B$ the magnetic field, $\mu$\footnote{$\mu = \mu_B/3$ for (CFL) color superconductivity and $\mu = \mu_B$ for proton superconductivity, where $\mu_B$ is the total baryon chemical potential.} is the chemical potential and $m$ is the mass of the HS, all defined at radius $r$\footnote{The calculations are done in $\hbar = c= 1$ units, with $\Delta_{\rm SC}$ and $\mu$ in MeV, and pressures expressed in MeV$^4$ units.}.

Here, the factors of $2m/r$ and $p_r/p_c$ ensure that the anisotropy profile follows the physical behavior such that it vanishes at the center and the surface of the star. This is important to ensure that our profiles follow the relevant stability criteria \cite{Herrera}. This is similar to the quasilocal model of anisotropy, where anisotropy is taken to be a function of the compactness, $2m/r$ \cite{QLaniso}. The contributions from magnetic field and superconductivity are in a functional form that match the order of their effects on the EoS, as this sets the pressure/energy scale. The signs `\,-\,' and `\,+\,' in the profile capture the possibility of RO and TO fields respectively, similar to Eq. \eqref{tov2}. As mentioned previously, in the current work, we consider TO fields in the star. This is consistent with our assumption of spherical symmetry \cite{pili} as well as stability arguments \cite{braithwaite}. Hence, we consider the `\,+\,' case in all the results reported presently.

An alternate scenario is the case where PT and/or the formation of superconducting phases in the HS can itself lead to pressure anisotropy in the star, independent of the magnetic field, as proposed in earlier work \cite{BL}. Superfluidity of neutrons has long been proposed to be one of the mechanisms responsible for pressure anisotropy \cite{superfluid_neutrons, superfluid_neutrons_2}. Further, certain types of superconductivity lead to anisotropic gaps (in the momentum space) which could lead to anisotropy at the macroscopic level. Additionally, there are also possible crystalline CSC phases \cite{ccs,ccs2} that could arise in the star, in which the deformation/shear is directly linked to the CSC gap.

To account for these possibilities, we propose an alternate profile (hereinafter, Profile 2), given by
\begin{equation}
\sigma_{\rm aniso} = p_t - p_r = \left(\pm B^2+\Delta_{\rm SC}^2\mu^2\right)\left(\frac{2m}{r}\right)\left(\frac{p_r}{p_c}\right).
\label{Xeqn7-3.2}
\end{equation}

Thus, in Profile 1, the role of superconductivity is \textit{multiplicative}, i.e., to supplement the magnetic stresses that lead to pressure anisotropy in the HS, while in Profile 2, either magnetic field or superconductivity could independently lead to anisotropy, and thus this shows the \textit{additive} contribution of superconductivity to the pressure anisotropy. We consider these two different models to capture the range of behaviors that superconductivity could have on the structures of NSs/HSs. Crucially, the profiles derived here are independent of the type of superconductivity present in the star. Naturally, the effect/role of proton superconductivity in such scenarios is lower when compared to quark cases due to the difference in the magnitude of their pairing gaps.

Since the exact microphysics of color superconductivity is unknown, we aim to parameterize the two possible facets of how superconductivity could affect the pressure/structure of the star through these two profiles. On one hand, in Profile 1, we have color superconductivity resulting in similar pressure modification as proton superconductivity \cite{type2_stress}. On the other hand, in Profile 2, we see color superconductivity leading to its own pressure anisotropy independent of the magnetic field. By comparing two profiles and their effects on the observables, we can isolate the effects of color superconductivity on the anisotropy and, hence, the structure of the HS. Based on observations, we also aim to constrain or rule out the model profile(s)/parameters.

These profiles are in contrast to other phenomenological models of anisotropy, e.g. the Bowers-Liang model \cite{BL}, where the parameterization is done in terms of an arbitrary parameter $\kappa$, with $\kappa = 0$ resulting in $\sigma_{\rm aniso} = 0$. In Profile 1, it is a physical quantity, the magnetic field $B$, that plays an equivalent role of $\kappa$. Further, since the effect of (proton/color) superconductivity in Profile 1 is modeled as following the physically established behavior of Type II proton superconductivity, when $B = 0$, $\sigma_{\rm aniso} = 0$. In Profile 2, we have expanded the role of $\kappa$ to be sourced by both $B$ and $\Delta_{\rm SC}$, such that when both $B = 0$ and $\Delta_{\rm SC} = 0$, then $\sigma_{\rm aniso} = 0$. This is due to the superconductivity itself leading to pressure anisotropy, following the behavior of, e.g., crystalline color superconductors.

As mentioned previously, the exact behavior of color superconductivity, and the quark phase in general, are largely unknown. In the current work, we have modeled the CSC quark core of HSs through a combination of quark EoS parameters, i.e. $K_v$, $B_{\rm eff}$ and $\Delta_{\rm CSC}$, and introduced phenomenological anisotropy profiles to capture the effects of the CSC quark phase on stellar structure. The profiles adopted here are not intended to be an exact microscopic description and, therefore, may carry modeling uncertainties. We explore a broad range of the quark matter parameters, particularly for $\Delta_{\rm CSC}$, to investigate the sensitivity of the stellar structure and anisotropic effects to the uncertain properties of the CSC phase. Further, note that our proposed anisotropic stresses arising from superconductivity are phenomenologically set equal to the superconducting condensation-energy scale
($\Delta_{\rm CSC}^{2}\mu^{2}$), up to an order-unity normalization, motivated by previous modeling of superconducting phases (e.g. \cite{ccs}). One may regard the uncertainties associated with any additional dimensionless coefficient as being effectively subsumed in the uncertainty of $\Delta_{\rm CSC}$. Thus, our results should be interpreted as a phenomenological exploration of the possible influence of magnetism and superconductivity rather than as a unique microscopic determination of the anisotropic stress.

The regions with proton superconductivity are identified by incorporating the analytical fitting for the superconducting gap $\Delta_{\rm SC}$ as a function of the number density within the star, given in previous work \cite{sinha_fit}. In the case of color superconductivity, we consider the entire quark sector to be color superconducting, with a constant gap throughout, $\Delta_{\rm CSC}$. The density dependence of the gap is not well established, owing to the uncertainties of color superconductivity. Constant $\Delta_{\rm CSC}$ is effectively valid in zero temperature systems like the ones we consider presently \cite{jaikumar_TB_gap}.

\section{Superconductivity zones}\label{Xsec5-4}
\label{sec:varySC}

\begin{figure*}[!htb]
    \centering
    \begin{subfigure}[t]{0.32\textwidth}
        \includegraphics[width=\textwidth]{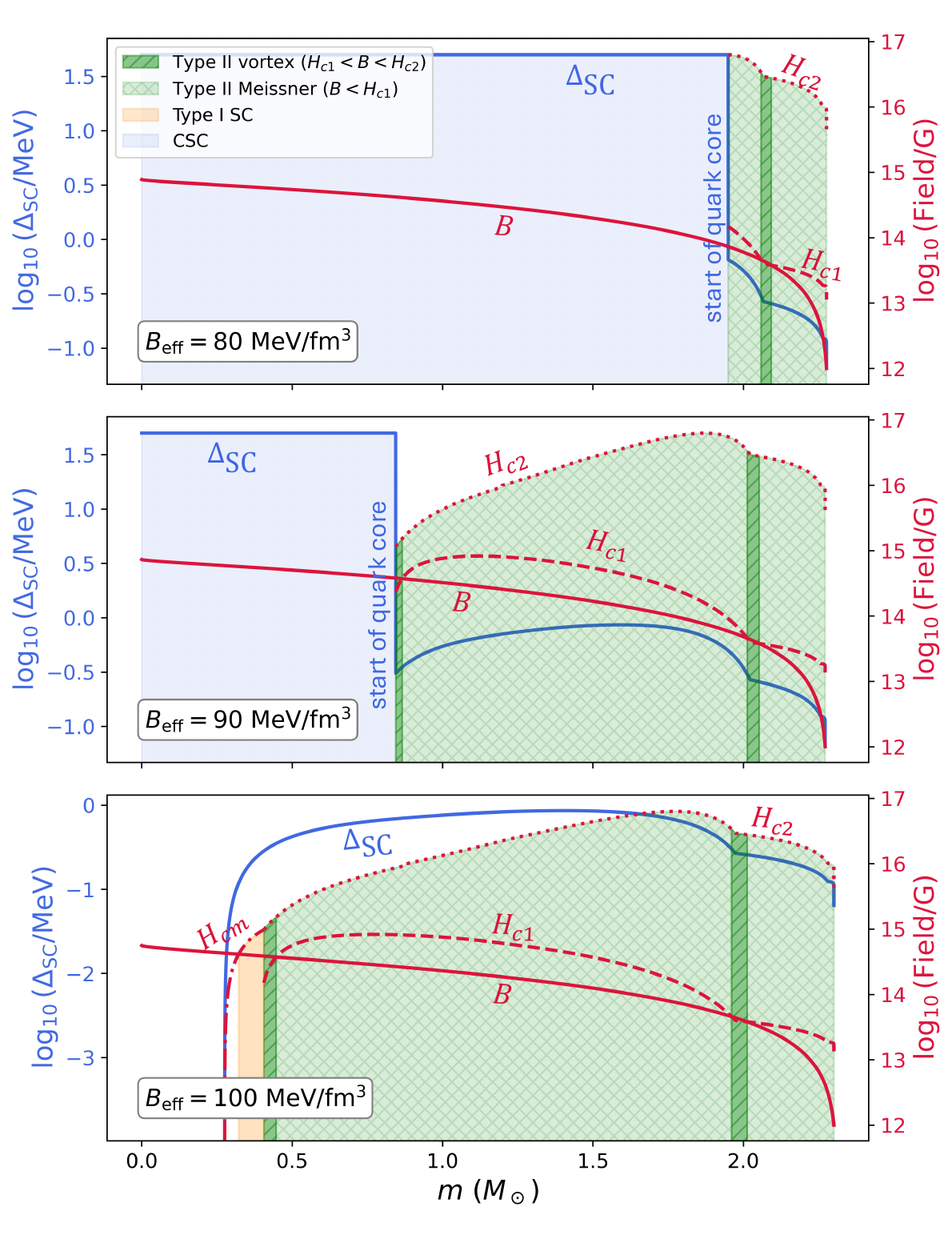}
    \end{subfigure}
    \hfill
    \begin{subfigure}[t]{0.32\textwidth}
        \includegraphics[width=\textwidth]{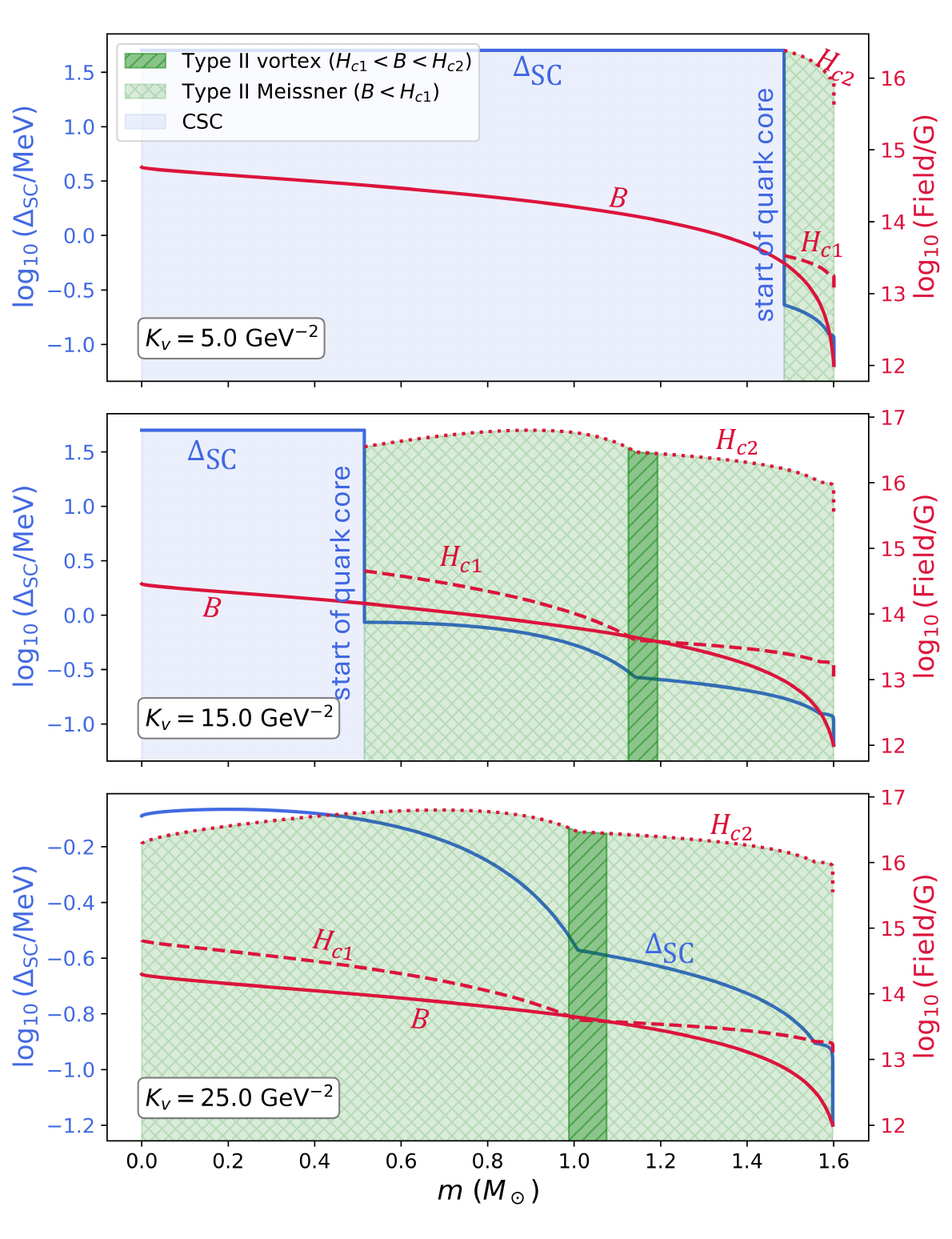}
    \end{subfigure}
    \hfill
    \begin{subfigure}[t]{0.32\textwidth}
        \includegraphics[width=\textwidth]{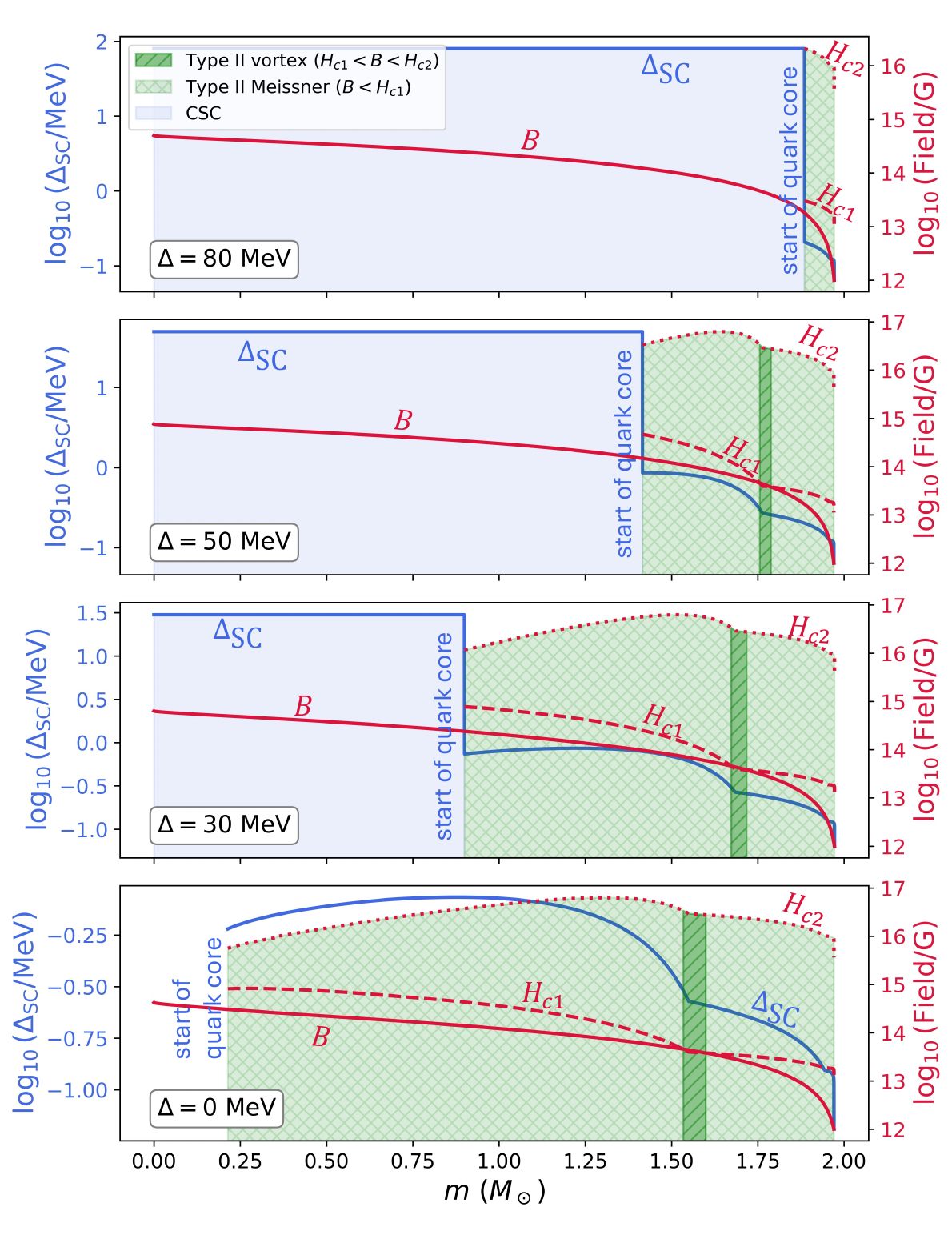}
    \end{subfigure}
    \caption{\justifying{Extent of superconductivity within a star (determined using the variation of superconducting gaps and fields as functions of stellar mass) with the change of parameters for a \textit{weakly} magnetized system - $B_0 = 10^{15} \ G, B_s = 10^{12} \ G$. \textit{Left}: $2.3M_\odot$ star for $\Delta_{\rm CSC} = 50 $ MeV, $K_v = 25$ GeV$^{-2}$, with varying $B_{\rm eff} = $ 80, 90, 100 MeV/fm$^3$. \textit{Middle}: $1.6M_\odot$ star for $\Delta_{\rm CSC} = 50 $ MeV, $B_{\rm eff} = 100$ MeV/fm$^{3}$, with varying $K_v = $ 5, 15, 25 GeV$^{-2}$. \textit{Right}: $2.1M_\odot$ star for $B_{\rm eff} = 100$ MeV/fm$^{3}$, $K_v = 15$ GeV$^{-2}$, with varying $\Delta_{\rm CSC} = $ 0, 30, 50, 80 MeV. The blue shading represents the CSC quark region. The light green shading (cross-hatched) represents Type II proton superconductivity in the Meissner state ($B < H_{c1}$). The dark green shading (dashed) represents Type II proton superconductivity in the vortex state ($H_{c1} < B < H_{c2}$). The orange shading represents the part of the star that is Type I proton superconducting ($B < H_{cm}$). }}
    \label{varySC_B015}
\end{figure*}

\begin{figure*}[!htb]
    \centering
    \begin{subfigure}[t]{0.32\textwidth}
        \includegraphics[width=\textwidth]{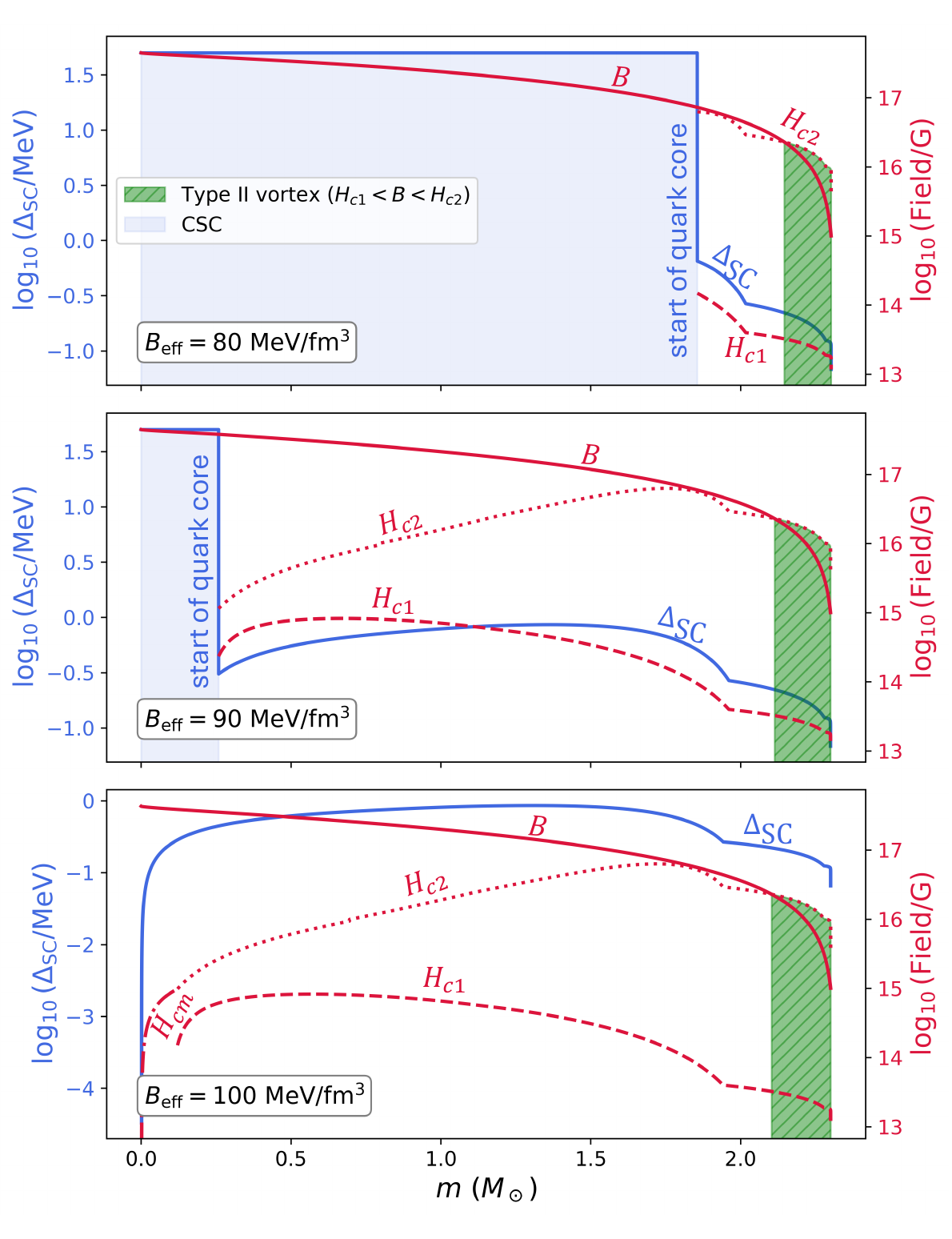}
    \end{subfigure}
    \hfill
    \begin{subfigure}[t]{0.32\textwidth}
        \includegraphics[width=\textwidth]{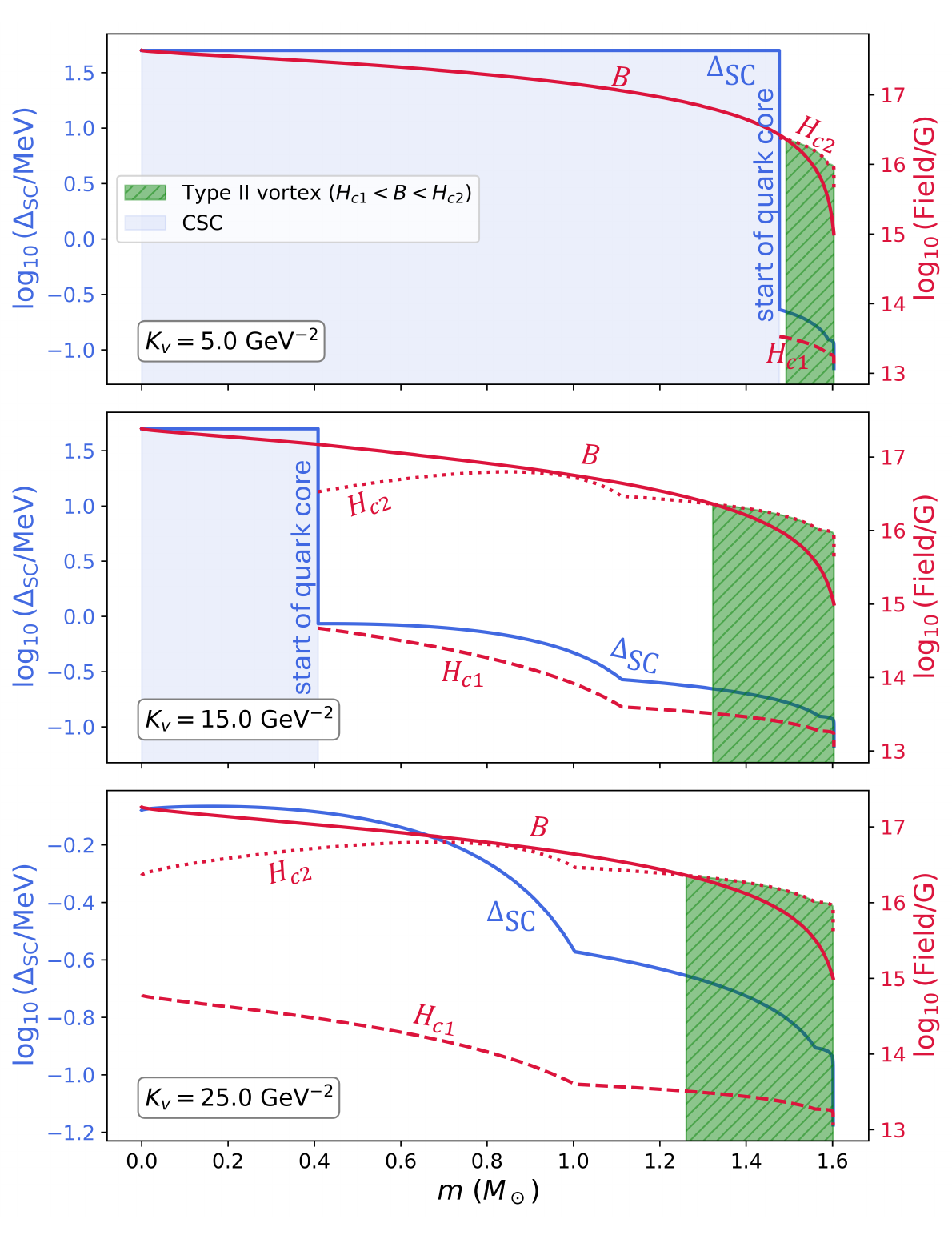}
    \end{subfigure}
    \hfill
    \begin{subfigure}[t]{0.32\textwidth}
        \includegraphics[width=\textwidth]{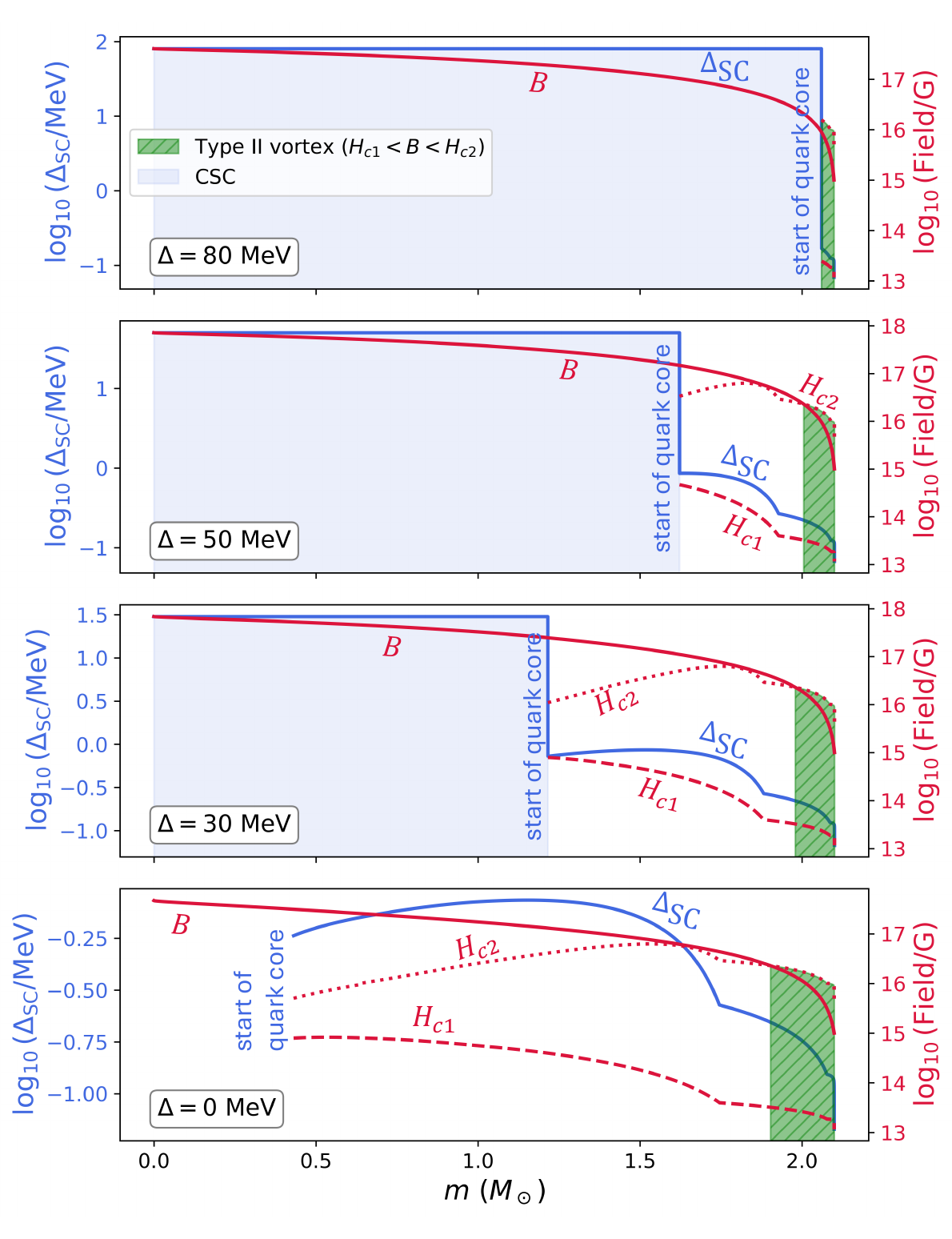}
    \end{subfigure}
    \caption{\justifying{Extent of superconductivity within a star with the change of parameters for a \textit{highly} magnetized system - $B_0 = 10^{18} \ G, B_s = 10^{15} \ G$. All parameter variations and shading are the same as Fig. \ref{varySC_B015}.}}
    \label{varySC_B018}
\end{figure*}

We are now all set to solve Eqs. (\ref{tov1}) and (\ref{tov2}) in order to obtain HSs with various properties.
A few representative cases for the variation of the superconducting pairing gap and superconducting regions within the star for different HSs are given in Figs. \ref{varySC_B015} and \ref{varySC_B018}. Each subfigure shows how changing the parameters - $K_v$, $B_{\rm eff}$ and $\Delta_{\rm CSC}$ - changes the extent and type of different superconducting regions within the star. The stellar structure is constructed with the contribution of superconducting anisotropy, specifically from Profile 2. The trends established in this section are general and follow for Profile 1 stars as well.

We find the following trends:
\begin{enumerate}
    \item An \textit{increase} in $B_{\rm eff}$ leads to the point of PT shifting to higher densities, and thus \textit{decreasing}  the extent of CSC matter in the star. A higher effective Bag constant refers to a higher energy cost to deconfine the quark phase, thus leading to higher pressures/densities required for the stable quark phase to appear in the star.
    \item An \textit{increase} in $K_v$ similarly shifts the point of PT to higher densities, and \textit{decreases} the extent of CSC matter in the star. Higher values of $K_v$ result in stiffer EoS in the quark phase, as now we have an additional vector repulsion contribution. This results in the pressure at a given $\mu$ reducing and, hence, we require higher pressures for the PT.
    \item An \textit{increase} in $\Delta_{\rm CSC}$ shifts the point of PT to lower densities, and \textit{increases} the extent of CSC matter in the star. This is due to the pressure and energy density being enhanced by the pairing energy $\sim \Delta_{\rm CSC}^2\mu^2$. This leads to the point of PT being reached at lower pressures.
\end{enumerate}

In summary, both $B_{\rm eff}$ and $K_v$ tend to decrease the extent of the quark core, with the increase in either of the quantities leading to higher pressures/densities required for the PT. The CSC gap has the opposite effect, as it promotes the quark core, while also softening the overall EoS. These trends also follow from the EoS trends noted in Fig. \ref{fig:EOS}.

The increase in magnetic field throughout the star affects mostly the proton superconducting phases in the hadronic part of the HS, as these phases are destroyed whenever the fields exceed the critical fields of superconductivity. The CSC phases are not affected as the critical fields for these cases are greater than $10^{18} \ G$. Proton superconductivity can be classified into Type I and Type II based on its response to the magnetic field. In Type II superconductivity, there are two critical fields, $H_{c1}$ and $H_{c2}$. For matter below $H_{c1}$, the superconducting matter expels all magnetic fields leading to a Meissner state. For fields between $H_{c1}$ and $H_{c2}$, the matter exists in a vortex state. Above $H_{c2}$, superconductivity is destroyed. The proton superconductivity in NSs is usually believed to be mostly Type II \cite{type2_ns, type2_ns2}. On the other hand, there are observations that indicate that there could be Type I superconductivity at play within NS cores \cite{type1_ns, type1_ns2}. In this case, there is one critical field $H_{cm}$, with Meissner expulsion below this field, and superconductivity absent above it.

The role of the magnetic field within the HS is then mostly to affect the proton superconducting zones. The quark EoS parameters, viz., $\Delta_{\rm CSC}, B_{\rm eff}$ and $K_v$, control the extent of the quark core, and hence the color superconductivity within the HS. The magnetic field does not affect the dynamics here \cite{jaikumar_TB_gap}. In the high magnetic field cases, there may be modification of  $\Delta_{\rm CSC}$ as the MCFL phase is favored over the CFL phase, particularly for low $\Delta_{\rm CSC}$ \cite{mcfl,mcfl_trans}. However, for the MCFL phase to significantly cause a difference in the EoS, $B \sim 10^{19} \ G$ is required, which is higher than the fields under consideration \cite{mcfl,LandauEOS,dc_mcfl}. Moreover, $B_{\rm eff}$ is unaffected for fields $<10^{20} \ G$, much higher than the plausible maximum field strengths for the systems under consideration \cite{mcfl}.  Hence, in the present work, the effect of the emergence of the MCFL phase is not practically important.

On the other hand, magnetic field can significantly affect the proton superconducting zones within the star. We see this when we compare the weakly magnetized cases in Fig. \ref{varySC_B015} (with $B_0 = 10^{15} \ G$) with the highly magnetized cases in Fig. \ref{varySC_B018} (with $B_0 = 10^{18} \ G$). In the high magnetic field cases, the proton superconductivity is confined to the outer layers of the star, corresponding to the crust. Any other superconductivity within the star comes about if a quark core arises in the star, bringing with it associated color superconductivity. In the low magnetic field cases, proton superconductivity itself can extend up to the point of PT, just before the appearance of the quark core and associated color superconductivity. Further, here, almost the entire proton superconducting zones are Type II Meissner, with limited appearance of Type II vortex and Type I states. It is important to note here that the magnetic flux expulsion associated with the Meissner states is not expected to significantly affect the magnetic fields of the star, as the flux expulsion timescale is comparable to the age of the Universe \cite{baym_meissner} (see, however: \cite{lander_meissner}, for an alternate mechanism of flux expulsion on shorter timescales). In the present work, we do not consider the dynamics of the magnetic field within the HS to be affected by either proton or color superconductivity.

Thus, an \textit{increase} in the magnetic fields throughout the HS leads to a \textit{decrease} in the extent of proton superconductivity in the star. This is the same reasoning used in previous work to rule out proton superconductivity in magnetar cores \cite{sinha_sc_magnetars,das_sc}. However, as we have shown using various combinations of parameters in Figs. \ref{varySC_B015} and \ref{varySC_B018}, superconductivity in the form of color superconductivity can arise in HS cores even for the high magnetic fields. Thus, the presence of quark cores in HSs can lead to the re-emergence of superconductivity, albeit in the form of color superconducting quarks.

\section{Effect on observables}\label{Xsec6-5}
\label{sec:obs}

\subsection{\label{Xsec7-5.1}
Effect on $M_{\rm {\max}}$}

Although the effect of color superconductivity on the EoS ($\simeq \Delta_{\rm CSC}^2\mu^2$) is expected to be sub-dominant when compared to the $\mu^4$ contribution from the Fermi sea, there are still situations where it can have a significant effect on the EoS. In particular, there may be regions in the star where the $\mu^4$ kinetic contribution cancels out with the Bag pressure, leading to the pairing energy of CSC matter to be the dominant term in the EoS \cite{reddy:CSC}. Further, the point of PT is also dependent on $\Delta_{\rm CSC}$, as seen in Fig. \ref{fig:EOS}. Thus, it is important and illustrative to include the effects of color superconductivity at the EoS level. We now examine in detail the structural effect of color superconductivity, particularly on the important observable that is $M_{\rm {\max}}$ of the star.

The presence of any pressure anisotropy generally brings associated mass enhancement. This is seen across different models for the anisotropy profile (e.g. \cite{zuraiq,aniso_HS_BL,aniso_HS_QL}). We have, in the past, examined the possibility of mass enhancement of NSs, using a combination of anisotropy and magnetic field, leading to masses in the lower mass gap, to explain objects like the secondary in GW190814 \cite{zuraiq}. We now examine if the structural effects on HSs from our current proposed anisotropy profiles can lead to similar conclusions.

Various $M_{\rm {\max}}$ obtained for various values of  $\Delta_{\rm CSC}$ and  $B_{\rm eff}$ are shown in Tables \ref{Table:prof1} and \ref{Table:prof2} for Profiles 1 and 2 respectively. The dash (``-'') entries represent cases where PT is not energetically favorable, and hence no HSs are formed. We use a fixed $K_v = 25 \ \rm{GeV}^{-2}$. We choose parameters such that the HS $M_{\rm {\max}} > 2M_\odot$, as constrained by observations. In these cases, the corresponding radii for $1.4 M_\odot$ star and $2M_\odot$ star are within $12.86-13.48$ km and $12.72 - 13.46$ km respectively, consistent with the most recent NICER constraints \cite{nicer1,nicer2}. In the case of stars with lower $K_v = 15 \ \rm{GeV}^{-2}$, the lower limit of radius is $11.59$ km and $11.22$ km for $1.4M_\odot$ and $2M_\odot$ stars respectively. Interestingly, stars with a higher fraction of quark matter/cores tend to have smaller radii, in line with the softening effects of quark matter.

\begin{table*}[]
\resizebox{\textwidth}{!}{\begin{tabular}{c|c|ccc|ccc|ccc|ccc}
\hline
\multirow{2}{*}{$\Delta_{\rm CSC}$ (MeV)} & \multirow{2}{*}{$B_0$ (G)} & \multicolumn{3}{c|}{$B_{\rm eff}$ = 120 MeV/fm$^3$}                                                                                   & \multicolumn{3}{c|}{$B_{\rm eff}$ = 100 MeV/fm$^3$}                                                                                   & \multicolumn{3}{c|}{$B_{\rm eff}$ = 90 MeV/fm$^3$}                                                                                    & \multicolumn{3}{c}{$B_{\rm eff}$ = 80 MeV/fm$^3$}                                                                                     \\ \cline{3-14} 
                                          &                            & \multicolumn{1}{c|}{$M_{\mathrm{max}} (M_\odot)$} & \multicolumn{1}{c|}{$\epsilon_c$ (MeV/fm$^3$)} & $\epsilon_{\mathrm{trans}}$ (MeV/fm$^3$) & \multicolumn{1}{c|}{$M_{\mathrm{max}} (M_\odot)$} & \multicolumn{1}{c|}{$\epsilon_c$ (MeV/fm$^3$)} & $\epsilon_{\mathrm{trans}}$ (MeV/fm$^3$) & \multicolumn{1}{c|}{$M_{\mathrm{max}} (M_\odot)$} & \multicolumn{1}{c|}{$\epsilon_c$ (MeV/fm$^3$)} & $\epsilon_{\mathrm{trans}}$ (MeV/fm$^3$) & \multicolumn{1}{c|}{$M_{\mathrm{max}} (M_\odot)$} & \multicolumn{1}{c|}{$\epsilon_c$ (MeV/fm$^3$)} & $\epsilon_{\mathrm{trans}}$ (MeV/fm$^3$) \\ \hline
\multirow{3}{*}{0}                        & Isotropic                  & \multicolumn{1}{c|}{2.442\textsuperscript{\textdagger}}                        & \multicolumn{1}{c|}{1055.33}               & 1055.33                              & \multicolumn{1}{c|}{2.432\textsuperscript{\textdagger}}                        & \multicolumn{1}{c|}{915.01}                & 915.01                               & \multicolumn{1}{c|}{2.412}                        & \multicolumn{1}{c|}{1027.27}               & 825.42                               & \multicolumn{1}{c|}{2.369}                        & \multicolumn{1}{c|}{956.02}                & 706.51                               \\
                                          & $10^{15}$                  & \multicolumn{1}{c|}{2.442\textsuperscript{\textdagger}}                        & \multicolumn{1}{c|}{1055.33}               & 1055.33                              & \multicolumn{1}{c|}{2.432\textsuperscript{\textdagger}}                        & \multicolumn{1}{c|}{915.01}                & 915.01                               & \multicolumn{1}{c|}{2.412}                        & \multicolumn{1}{c|}{1027.27}               & 825.42                               & \multicolumn{1}{c|}{2.369}                        & \multicolumn{1}{c|}{956.02}                & 706.51                               \\
                                          & $10^{18}$                  & \multicolumn{1}{c|}{2.562\textsuperscript{\textdagger}}                        & \multicolumn{1}{c|}{887.48}                & 1055.33                              & \multicolumn{1}{c|}{2.562\textsuperscript{\textdagger}}                        & \multicolumn{1}{c|}{887.48}                & 915.01                               & \multicolumn{1}{c|}{2.558\textsuperscript{\textdagger}}                        & \multicolumn{1}{c|}{825.42}                & 825.42                               & \multicolumn{1}{c|}{2.518}                        & \multicolumn{1}{c|}{822.05}                & 706.51                               \\ \hline
\multirow{3}{*}{10}                       & Isotropic                  & \multicolumn{1}{c|}{2.442\textsuperscript{\textdagger}}                        & \multicolumn{1}{c|}{1042.32}               & 1042.32                              & \multicolumn{1}{c|}{2.43\textsuperscript{\textdagger}}                         & \multicolumn{1}{c|}{906.36}                & 906.36                               & \multicolumn{1}{c|}{2.408}                        & \multicolumn{1}{c|}{1014.67}               & 812.30                               & \multicolumn{1}{c|}{2.365}                        & \multicolumn{1}{c|}{953.85}                & 694.36                               \\
                                          & $10^{15}$                  & \multicolumn{1}{c|}{2.442\textsuperscript{\textdagger}}                        & \multicolumn{1}{c|}{1042.32}               & 1042.32                              & \multicolumn{1}{c|}{2.43\textsuperscript{\textdagger}}                         & \multicolumn{1}{c|}{906.36}                & 906.36                               & \multicolumn{1}{c|}{2.408}                        & \multicolumn{1}{c|}{1014.67}               & 812.30                               & \multicolumn{1}{c|}{2.365}                        & \multicolumn{1}{c|}{953.85}                & 694.36                               \\
                                          & $10^{18}$                  & \multicolumn{1}{c|}{2.562\textsuperscript{\textdagger}}                        & \multicolumn{1}{c|}{887.48}                & 1042.32                              & \multicolumn{1}{c|}{2.562\textsuperscript{\textdagger}}                        & \multicolumn{1}{c|}{887.48}                & 906.36                               & \multicolumn{1}{c|}{2.556\textsuperscript{\textdagger}}                        & \multicolumn{1}{c|}{812.30}                & 812.30                               & \multicolumn{1}{c|}{2.511}                        & \multicolumn{1}{c|}{810.98}                & 694.36                               \\ \hline
\multirow{3}{*}{30}                       & Isotropic                  & \multicolumn{1}{c|}{2.438\textsuperscript{\textdagger}}                        & \multicolumn{1}{c|}{971.35}                & 971.35                               & \multicolumn{1}{c|}{2.411}                        & \multicolumn{1}{c|}{1030.94}               & 818.85                               & \multicolumn{1}{c|}{2.368}                        & \multicolumn{1}{c|}{943.25}                & 712.62                               & \multicolumn{1}{c|}{2.302}                        & \multicolumn{1}{c|}{1010.94}               & 538.95                               \\
                                          & $10^{15}$                  & \multicolumn{1}{c|}{2.438\textsuperscript{\textdagger}}                        & \multicolumn{1}{c|}{971.35}                & 971.35                               & \multicolumn{1}{c|}{2.411\textsuperscript{\textdagger}}                        & \multicolumn{1}{c|}{1030.94}               & 818.85                               & \multicolumn{1}{c|}{2.368}                        & \multicolumn{1}{c|}{943.25}                & 712.62                               & \multicolumn{1}{c|}{2.302}                        & \multicolumn{1}{c|}{1010.94}               & 538.95                               \\
                                          & $10^{18}$                  & \multicolumn{1}{c|}{2.562\textsuperscript{\textdagger}}                        & \multicolumn{1}{c|}{887.48}                & 971.35                               & \multicolumn{1}{c|}{2.557\textsuperscript{\textdagger}}                        & \multicolumn{1}{c|}{818.85}                & 818.85                               & \multicolumn{1}{c|}{2.521}                        & \multicolumn{1}{c|}{844.52}                & 712.62                               & \multicolumn{1}{c|}{2.433}                        & \multicolumn{1}{c|}{868.77}                & 538.95                               \\ \hline
\multirow{3}{*}{50}                       & Isotropic                  & \multicolumn{1}{c|}{2.411}                        & \multicolumn{1}{c|}{1070.35}               & 825.42                               & \multicolumn{1}{c|}{2.295\textsuperscript{\textdagger}}                        & \multicolumn{1}{c|}{640.97}                & 640.97                               & \multicolumn{1}{c|}{2.253}                        & \multicolumn{1}{c|}{1050.92}               & 486.49                               & \multicolumn{1}{c|}{2.248}                        & \multicolumn{1}{c|}{1131.53}               & 245.04                               \\
                                          & $10^{15}$                  & \multicolumn{1}{c|}{2.411}                        & \multicolumn{1}{c|}{1070.35}               & 825.42                               & \multicolumn{1}{c|}{2.295\textsuperscript{\textdagger}}                        & \multicolumn{1}{c|}{640.97}                & 640.97                               & \multicolumn{1}{c|}{2.253}                        & \multicolumn{1}{c|}{1050.92}               & 486.49                               & \multicolumn{1}{c|}{2.248}                        & \multicolumn{1}{c|}{1131.53}               & 245.04                               \\
                                          & $10^{18}$                  & \multicolumn{1}{c|}{2.558\textsuperscript{\textdagger}}                        & \multicolumn{1}{c|}{825.42}                & 825.42                               & \multicolumn{1}{c|}{2.461}                        & \multicolumn{1}{c|}{814.75}                & 640.97                               & \multicolumn{1}{c|}{2.367}                        & \multicolumn{1}{c|}{921.89}                & 486.49                               & \multicolumn{1}{c|}{2.437}                        & \multicolumn{1}{c|}{873.30}                & 245.04                               \\ \hline
\multirow{3}{*}{80}                       & Isotropic                  & \multicolumn{1}{c|}{2.179}                        & \multicolumn{1}{c|}{1046.44}               & 505.88                               & \multicolumn{1}{c|}{2.151}                        & \multicolumn{1}{c|}{1249.07}               & 271.02                               & \multicolumn{1}{c|}{-}                            & \multicolumn{1}{c|}{-}                     & -                                    & \multicolumn{1}{c|}{-}                            & \multicolumn{1}{c|}{-}                     & -                                    \\
                                          & $10^{15}$                  & \multicolumn{1}{c|}{2.179}                        & \multicolumn{1}{c|}{1067.50}               & 505.88                               & \multicolumn{1}{c|}{2.151}                        & \multicolumn{1}{c|}{1249.07}               & 271.02                               & \multicolumn{1}{c|}{-}                            & \multicolumn{1}{c|}{-}                     & -                                    & \multicolumn{1}{c|}{-}                            & \multicolumn{1}{c|}{-}                     & -                                    \\
                                          & $10^{18}$                  & \multicolumn{1}{c|}{2.385}                        & \multicolumn{1}{c|}{887.48}                & 505.88                               & \multicolumn{1}{c|}{2.445}                        & \multicolumn{1}{c|}{848.80}                & 271.02                               & \multicolumn{1}{c|}{-}                            & \multicolumn{1}{c|}{-}                     & -                                    & \multicolumn{1}{c|}{-}                            & \multicolumn{1}{c|}{-}                     & -                                    \\ \hline
\multirow{3}{*}{100}                      & Isotropic                  & \multicolumn{1}{c|}{2.068}                        & \multicolumn{1}{c|}{1374.99}               & 294.37                               & \multicolumn{1}{c|}{-}                            & \multicolumn{1}{c|}{-}                     & -                                    & \multicolumn{1}{c|}{-}                            & \multicolumn{1}{c|}{-}                     & -                                    & \multicolumn{1}{c|}{-}                            & \multicolumn{1}{c|}{-}                     & -                                    \\
                                          & $10^{15}$                  & \multicolumn{1}{c|}{2.068}                        & \multicolumn{1}{c|}{1374.99}               & 294.37                               & \multicolumn{1}{c|}{-}                            & \multicolumn{1}{c|}{-}                     & -                                    & \multicolumn{1}{c|}{-}                            & \multicolumn{1}{c|}{-}                     & -                                    & \multicolumn{1}{c|}{-}                            & \multicolumn{1}{c|}{-}                     & -                                    \\
                                          & $10^{18}$                  & \multicolumn{1}{c|}{2.476}                        & \multicolumn{1}{c|}{820.78}                & 294.37                               & \multicolumn{1}{c|}{-}                            & \multicolumn{1}{c|}{-}                     & -                                    & \multicolumn{1}{c|}{-}                            & \multicolumn{1}{c|}{-}                     & -                                    & \multicolumn{1}{c|}{-}                            & \multicolumn{1}{c|}{-}                     & -                                    \\ \hline
\end{tabular}}
\caption{\justifying{$M_{\rm max}$ for various $B_{\rm eff}$ and $\Delta_{\rm CSC}$, as calculated for Profile 1 anisotropy. The corresponding central density ($\epsilon_c$) and density of hadron matter at the PT ($\epsilon_{\mathrm{trans}}$) for each case is also given.} In each case, the results are shown for the Isotropic case: $\sigma_{\rm aniso} = 0$; weakly magnetized case: $B_0 = 10^{15} \ G, B_s = 10^{12} \ G$; and highly magnetized case: $B_0 = 10^{18} \ G, B_s = 10^{15} \ G$.  The symbol `{\textdagger}' indicates the cases where the deconfinement PT leads to an unstable branch ($\epsilon_c \leq \epsilon_{\mathrm{trans}}$) and thus the $M_{\mathrm{max}}$ corresponds to a hadronic star.}
\label{Table:prof1}
\end{table*}

\begin{table*}[]
\resizebox{\textwidth}{!}{\begin{tabular}{c|c|ccc|ccc|ccc|ccc}
\hline
\multirow{2}{*}{$\Delta_{\rm CSC}$ (MeV)} & \multirow{2}{*}{$B_0$ (G)} & \multicolumn{3}{c|}{$B_{\rm eff}$ = 120 MeV/fm$^3$}                                                                                   & \multicolumn{3}{c|}{$B_{\rm eff}$ = 100 MeV/fm$^3$}                                                                                   & \multicolumn{3}{c|}{$B_{\rm eff}$ = 90 MeV/fm$^3$}                                                                                    & \multicolumn{3}{c}{$B_{\rm eff}$ = 80 MeV/fm$^3$}                                                                                     \\ \cline{3-14} 
                                          &                            & \multicolumn{1}{c|}{$M_{\mathrm{max}} (M_\odot)$} & \multicolumn{1}{c|}{$\epsilon_c$ (MeV/fm$^3$)} & $\epsilon_{\mathrm{trans}}$ (MeV/fm$^3$) & \multicolumn{1}{c|}{$M_{\mathrm{max}} (M_\odot)$} & \multicolumn{1}{c|}{$\epsilon_c$ (MeV/fm$^3$)} & $\epsilon_{\mathrm{trans}}$ (MeV/fm$^3$) & \multicolumn{1}{c|}{$M_{\mathrm{max}} (M_\odot)$} & \multicolumn{1}{c|}{$\epsilon_c$ (MeV/fm$^3$)} & $\epsilon_{\mathrm{trans}}$ (MeV/fm$^3$) & \multicolumn{1}{c|}{$M_{\mathrm{max}} (M_\odot)$} & \multicolumn{1}{c|}{$\epsilon_c$ (MeV/fm$^3$)} & $\epsilon_{\mathrm{trans}}$ (MeV/fm$^3$) \\ \hline
\multirow{3}{*}{0}                        & Isotropic                  & \multicolumn{1}{c|}{2.442\textsuperscript{\textdagger}}                        & \multicolumn{1}{c|}{1055.33}               & 1055.33                              & \multicolumn{1}{c|}{2.432\textsuperscript{\textdagger}}                        & \multicolumn{1}{c|}{915.01}                & 915.01                               & \multicolumn{1}{c|}{2.412}                        & \multicolumn{1}{c|}{1027.27}               & 825.42                               & \multicolumn{1}{c|}{2.369}                        & \multicolumn{1}{c|}{956.02}                & 706.51                               \\
                                          & $10^{15}$                  & \multicolumn{1}{c|}{2.442\textsuperscript{\textdagger}}                        & \multicolumn{1}{c|}{1055.33}               & 1055.33                              & \multicolumn{1}{c|}{2.432\textsuperscript{\textdagger}}                        & \multicolumn{1}{c|}{915.01}                & 915.01                               & \multicolumn{1}{c|}{2.412}                        & \multicolumn{1}{c|}{1027.27}               & 825.42                               & \multicolumn{1}{c|}{2.369}                        & \multicolumn{1}{c|}{956.02}                & 706.51                               \\
                                          & $10^{18}$                  & \multicolumn{1}{c|}{2.562\textsuperscript{\textdagger}}                        & \multicolumn{1}{c|}{887.48}                & 1055.33                              & \multicolumn{1}{c|}{2.562\textsuperscript{\textdagger}}                        & \multicolumn{1}{c|}{887.48}                & 915.01                               & \multicolumn{1}{c|}{2.558\textsuperscript{\textdagger}}                        & \multicolumn{1}{c|}{825.42}                & 825.42                               & \multicolumn{1}{c|}{2.518}                        & \multicolumn{1}{c|}{822.05}                & 706.51                               \\ \hline
\multirow{3}{*}{10}                       & Isotropic                  & \multicolumn{1}{c|}{2.442\textsuperscript{\textdagger}}                        & \multicolumn{1}{c|}{1042.32}               & 1042.32                              & \multicolumn{1}{c|}{2.43\textsuperscript{\textdagger}}                         & \multicolumn{1}{c|}{906.36}                & 906.36                               & \multicolumn{1}{c|}{2.408}                        & \multicolumn{1}{c|}{1014.67}               & 812.30                               & \multicolumn{1}{c|}{2.365}                        & \multicolumn{1}{c|}{953.85}                & 694.36                               \\
                                          & $10^{15}$                  & \multicolumn{1}{c|}{2.442\textsuperscript{\textdagger}}                        & \multicolumn{1}{c|}{1042.32}               & 1042.32                              & \multicolumn{1}{c|}{2.43}                         & \multicolumn{1}{c|}{1142.17}               & 906.36                               & \multicolumn{1}{c|}{2.408}                        & \multicolumn{1}{c|}{1014.67}               & 812.30                               & \multicolumn{1}{c|}{2.365}                        & \multicolumn{1}{c|}{953.85}                & 694.36                               \\
                                          & $10^{18}$                  & \multicolumn{1}{c|}{2.562\textsuperscript{\textdagger}}                        & \multicolumn{1}{c|}{887.48}                & 1042.32                              & \multicolumn{1}{c|}{2.562\textsuperscript{\textdagger}}                        & \multicolumn{1}{c|}{887.48}                & 906.36                               & \multicolumn{1}{c|}{2.556\textsuperscript{\textdagger}}                        & \multicolumn{1}{c|}{812.30}                & 812.30                               & \multicolumn{1}{c|}{2.511}                        & \multicolumn{1}{c|}{810.98}                & 694.36                               \\ \hline
\multirow{3}{*}{30}                       & Isotropic                  & \multicolumn{1}{c|}{2.438\textsuperscript{\textdagger}}                        & \multicolumn{1}{c|}{971.35}                & 971.35                               & \multicolumn{1}{c|}{2.411}                        & \multicolumn{1}{c|}{1030.94}               & 818.85                               & \multicolumn{1}{c|}{2.368}                        & \multicolumn{1}{c|}{943.25}                & 712.62                               & \multicolumn{1}{c|}{2.302}                        & \multicolumn{1}{c|}{1010.94}               & 538.95                               \\
                                          & $10^{15}$                  & \multicolumn{1}{c|}{2.438\textsuperscript{\textdagger}}                        & \multicolumn{1}{c|}{971.35}                & 971.35                               & \multicolumn{1}{c|}{2.411}                        & \multicolumn{1}{c|}{1030.94}               & 818.85                               & \multicolumn{1}{c|}{2.369}                        & \multicolumn{1}{c|}{946.55}                & 712.62                               & \multicolumn{1}{c|}{2.310}                        & \multicolumn{1}{c|}{1000.59}               & 538.95                               \\
                                          & $10^{18}$                  & \multicolumn{1}{c|}{2.562\textsuperscript{\textdagger}}                        & \multicolumn{1}{c|}{887.48}                & 971.35                               & \multicolumn{1}{c|}{2.557\textsuperscript{\textdagger}}                        & \multicolumn{1}{c|}{818.85}                & 818.85                               & \multicolumn{1}{c|}{2.521}                        & \multicolumn{1}{c|}{844.52}                & 712.62                               & \multicolumn{1}{c|}{2.434}                        & \multicolumn{1}{c|}{868.77}                & 538.95                               \\ \hline
\multirow{3}{*}{50}                       & Isotropic                  & \multicolumn{1}{c|}{2.411}                        & \multicolumn{1}{c|}{1070.35}               & 825.42                               & \multicolumn{1}{c|}{2.295\textsuperscript{\textdagger}}                        & \multicolumn{1}{c|}{640.97}                & 640.97                               & \multicolumn{1}{c|}{2.253}                        & \multicolumn{1}{c|}{1050.92}               & 486.49                               & \multicolumn{1}{c|}{2.248}                        & \multicolumn{1}{c|}{1131.53}               & 245.04                               \\
                                          & $10^{15}$                  & \multicolumn{1}{c|}{2.411}                        & \multicolumn{1}{c|}{1066.85}               & 825.42                               & \multicolumn{1}{c|}{2.325}                        & \multicolumn{1}{c|}{935.14}                & 640.97                               & \multicolumn{1}{c|}{2.282}                        & \multicolumn{1}{c|}{1012.50}               & 486.49                               & \multicolumn{1}{c|}{2.303}                        & \multicolumn{1}{c|}{999.05}                & 245.04                               \\
                                          & $10^{18}$                  & \multicolumn{1}{c|}{2.558\textsuperscript{\textdagger}}                        & \multicolumn{1}{c|}{825.42}                & 825.42                               & \multicolumn{1}{c|}{2.470}                        & \multicolumn{1}{c|}{795.44}                & 640.97                               & \multicolumn{1}{c|}{2.414}                        & \multicolumn{1}{c|}{870.67}                & 486.49                               & \multicolumn{1}{c|}{2.455}                        & \multicolumn{1}{c|}{841.94}                & 245.04                               \\ \hline
\multirow{3}{*}{80}                       & Isotropic                  & \multicolumn{1}{c|}{2.179}                        & \multicolumn{1}{c|}{1046.44}               & 505.88                               & \multicolumn{1}{c|}{2.151}                        & \multicolumn{1}{c|}{1249.07}               & 271.02                               & \multicolumn{1}{c|}{-}                            & \multicolumn{1}{c|}{-}                     & -                                    & \multicolumn{1}{c|}{-}                            & \multicolumn{1}{c|}{-}                     & -                                    \\
                                          & $10^{15}$                  & \multicolumn{1}{c|}{2.241}                        & \multicolumn{1}{c|}{988.21}                & 505.88                               & \multicolumn{1}{c|}{2.291}                        & \multicolumn{1}{c|}{905.64}                & 271.02                               & \multicolumn{1}{c|}{-}                            & \multicolumn{1}{c|}{-}                     & -                                    & \multicolumn{1}{c|}{-}                            & \multicolumn{1}{c|}{-}                     & -                                    \\
                                          & $10^{18}$                  & \multicolumn{1}{c|}{2.396}                        & \multicolumn{1}{c|}{862.63}                & 505.88                               & \multicolumn{1}{c|}{2.492}                        & \multicolumn{1}{c|}{770.84}                & 271.02                               & \multicolumn{1}{c|}{-}                            & \multicolumn{1}{c|}{-}                     & -                                    & \multicolumn{1}{c|}{-}                            & \multicolumn{1}{c|}{-}                     & -                                    \\ \hline
\multirow{3}{*}{100}                      & Isotropic                  & \multicolumn{1}{c|}{2.068}                        & \multicolumn{1}{c|}{1374.99}               & 294.37                               & \multicolumn{1}{c|}{-}                            & \multicolumn{1}{c|}{-}                     & -                                    & \multicolumn{1}{c|}{-}                            & \multicolumn{1}{c|}{-}                     & -                                    & \multicolumn{1}{c|}{-}                            & \multicolumn{1}{c|}{-}                     & -                                    \\
                                          & $10^{15}$                  & \multicolumn{1}{c|}{2.291}                        & \multicolumn{1}{c|}{811.82}                & 294.37                               & \multicolumn{1}{c|}{-}                            & \multicolumn{1}{c|}{-}                     & -                                    & \multicolumn{1}{c|}{-}                            & \multicolumn{1}{c|}{-}                     & -                                    & \multicolumn{1}{c|}{-}                            & \multicolumn{1}{c|}{-}                     & -                                    \\
                                          & $10^{18}$                  & \multicolumn{1}{c|}{2.568}                        & \multicolumn{1}{c|}{680.05}                & 294.37                               & \multicolumn{1}{c|}{-}                            & \multicolumn{1}{c|}{-}                     & -                                    & \multicolumn{1}{c|}{-}                            & \multicolumn{1}{c|}{-}                     & -                                    & \multicolumn{1}{c|}{-}                            & \multicolumn{1}{c|}{-}                     & -                                    \\ \hline
\end{tabular}}
\caption{\justifying{Same as Table \ref{Table:prof1}, except for Profile 2 anisotropy.}}
\label{Table:prof2}
\end{table*}

As noted in the previous section and from the trends of EoS/point of PT, larger pairing energy of color superconductivity in the star, i.e., due to a larger value of $\Delta_{\rm CSC}$, leads to shifting of PT to lower densities and thus the star is populated with more quark matter. This effect leads to the softer quark EoS occupying more of the star and, hence, $M_{\rm {\max}}$ goes down with an increase of $\Delta_{\rm CSC}$. This is illustrated from the ``Isotropic'' columns of Tables \ref{Table:prof1} and \ref{Table:prof2}. This is a ``color superconductivity softening'', analogous to the well-known problem of hyperon softening  \cite{hyperons} discussed in the NS literature. We find that effect of the color superconductivity softening is dominant over the anisotropic effect for $\Delta_{\rm CSC} < 50$ MeV.

However, in our models, the role of superconductivity extends beyond influencing the point of PT, and manifests in the pressure anisotropy. We now examine how the $M_{\rm {\max}}$ further changes in the two regimes of anisotropy we consider - Profile 1: where the role of $\Delta_{\rm SC}$ is to supplement the magnetic stresses, and Profile 2: where $\Delta_{\rm SC}$ has its own contribution to the pressure anisotropic stresses.

In the presence of high fields, i.e., $B_0 = 10^{18}$ G (corresponding to ``magnetar'' type systems: $B_s = 10^{15} \ G$), the magnetic field and anisotropy both contribute to the overall pressure anisotropy. This corresponds to maximum central/internal fields $B_c \simeq 9.5 \times 10^{17} \ G$, depending on the density, with magnetic-to-gravitational energy ratio $\approx 0.03$, ensuring our stars are stable \cite{braithwaite}. We find that although there is color superconductivity softening, the presence of magnetic fields and pressure anisotropy can help the star to still reach high masses of $2M_\odot$ and higher (even up to $2.5M_\odot$). This is similar to the hyperon admixed mass gap candidates in the presence of magnetic fields we studied in previous work \cite{zuraiq}. When comparing across Profile 1 with Profile 2, we find that Profile 1 gives slightly lower $M_{\rm {\max}}$, with the effect mostly pronounced for high gaps ($\Delta_{\rm CSC} > 50$ MeV).

\begin{figure}[!htb]
     \centering
         {\centering
         \includegraphics[width=0.5\textwidth]{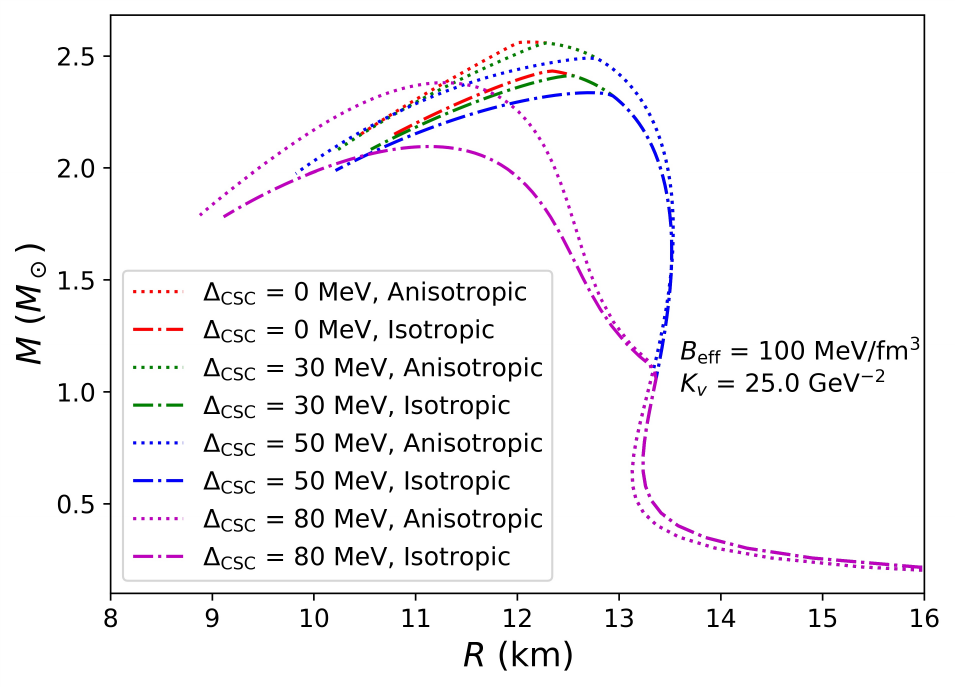}

     \vskip\baselineskip
            \includegraphics[width=0.5\textwidth]{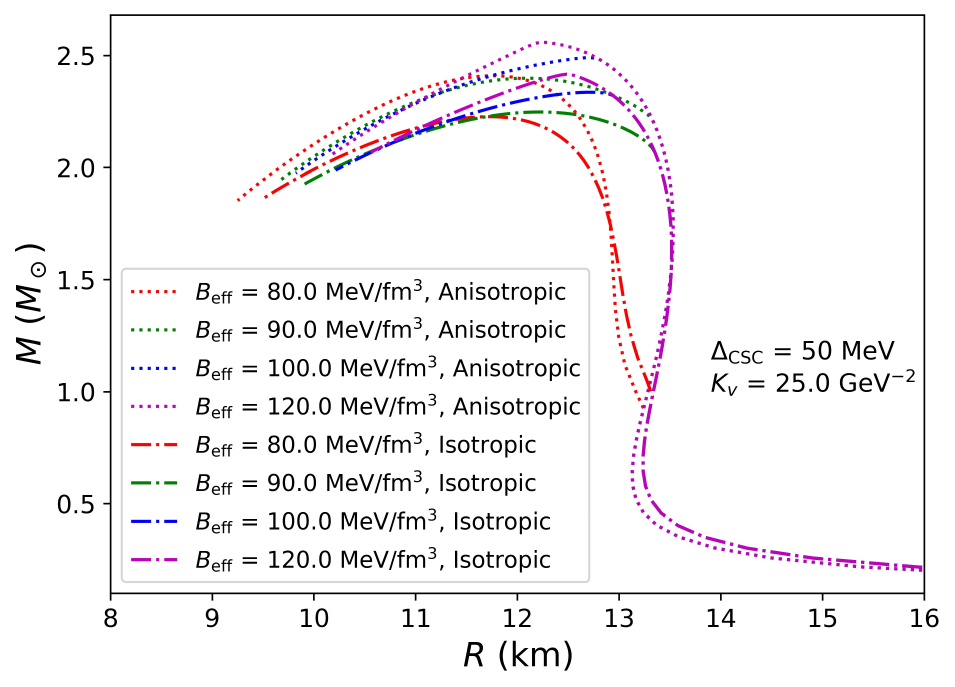}}

\caption{\justifying{\textit{Upper: }$M-R$ curves for different values of $\Delta_{\rm CSC}$ with Profile 1 anisotropy. Dotted lines indicate anisotropic stars. Dot-dashed lines are isotropic stars. $B_{\rm eff} = 100$ MeV/fm$^3$, $K_v = 25$ GeV$^{-2}$. \textit{Lower: } Variation of $M-R$ curves with different values of $B_{\rm eff}$. Here, $\Delta_{\rm CSC} = 50$ MeV, $K_v = 25$ GeV$^{-2}$. In all cases, we consider ``TO" fields with $B_0 = 10^{18} \ G, B_s = 10^{15} \ G$, $\eta = 0.01, \gamma = 1$. }}
    \label{MR_prof1}
\end{figure}

\begin{figure}[!htb]
     \centering
         {\centering
         \includegraphics[width=0.5\textwidth]{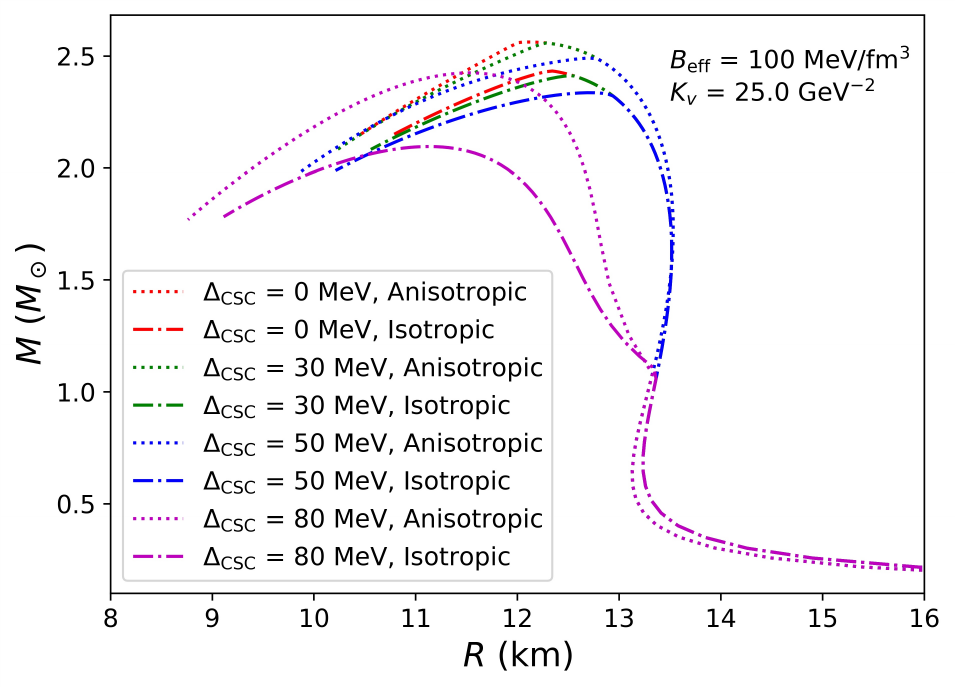}

     \vskip\baselineskip
            \includegraphics[width=0.5\textwidth]{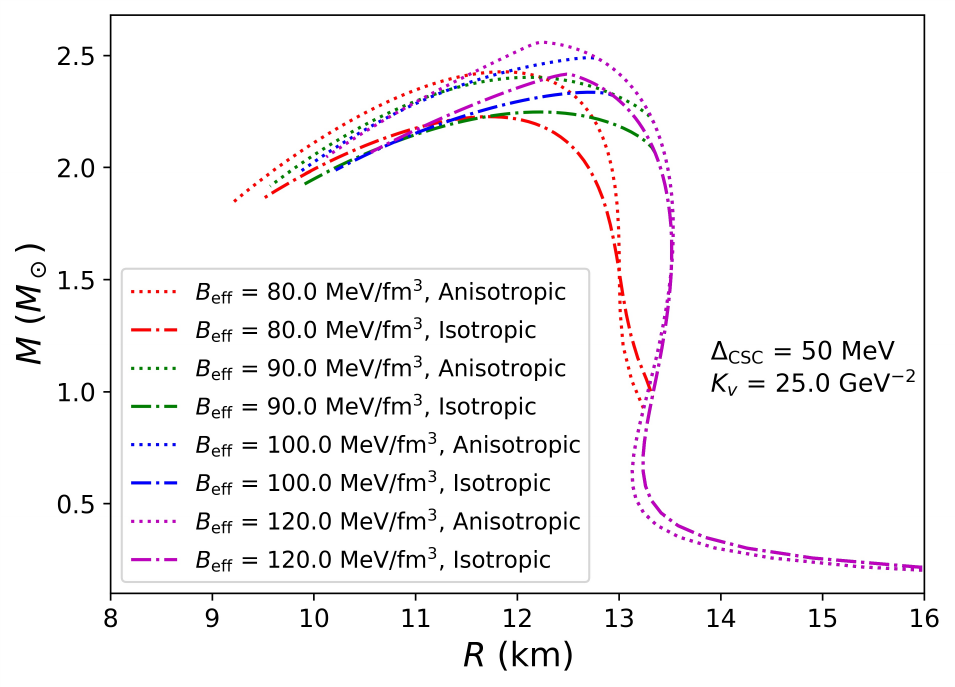}}

\caption{\justifying{Same as Fig. \ref{MR_prof1}, except with Profile 2 anisotropy.}}
    \label{MR_prof2}
\end{figure}

We can further examine the effect of magnetic fields on the $M-R$ curves, as shown in Figs. \ref{MR_prof1} and \ref{MR_prof2}, respectively, for Profile 1 and Profile 2. One important aspect to consider is that the PT need not necessarily take the star to a stable HS branch. Instead, if the point of PT is at high densities, the PT may lead to an unstable branch. These cases can be singled out by comparing the central density ($\epsilon_c$) of the $M_{\mathrm{max}}$ star with the density of hadron matter at the point of PT, $\epsilon_{\mathrm{trans}}$, as done in Tables  \ref{Table:prof1} and \ref{Table:prof2}. If $\epsilon_c < \epsilon_{\mathrm{trans}}$, a stable configuration containing a quark core is not attained, i.e., the HS branch is unstable, since $dM/d\epsilon_c < 0$. In these cases, the role of anisotropy is to enhance the mass of the hadron star before it undergoes PT to an unstable branch. A similar effect was previously observed for different models of HS EoS and anisotropy~\cite{zz_univ}. Thus, particularly in the cases where PT happens at high masses ($> 2M_\odot$), the role of the high magnetic field is to enable the star to attain more mass in the hadronic phase before it undergoes a PT. In the present model, the combination of magnetic field and anisotropy enables the NS to reach higher masses before undergoing a PT, and, thus, the hadronic matter in the star itself can support masses up to the lower mass gap ($\approx 2.5 M_\odot$ and higher).

On the other hand, in the presence of low fields, i.e., $B_0 = 10^{15}$ G  (corresponding to regular NS type systems: $B_s = 10^{12} \ G$), it is $\Delta_{\rm CSC}$ which plays the central role in determining $M_{\mathrm{max}}$. The difference between the two profiles is more distinct in this case, as seen from Tables \ref{Table:prof1} and \ref{Table:prof2}. This is because in Profile 1, the effective pressure anisotropy due to color superconductivity is directly tied to the magnetic field and its effects, while in Profile 2, the CSC gap can give rise to its own anisotropy. Comparing with the most extreme case, i.e., $B_{\rm eff} = 120$ MeV/fm$^{3}$ and $\Delta_{\rm CSC} = 100$ MeV, we find that the isotropic $M_{\mathrm{max}}$ gets enhanced to $2.291M_\odot$ in Profile 2 from $2.068M_\odot$, while there is no enhancement in Profile 1. Once again, for CSC gap to play an important structural role in terms of anisotropy, we require $\Delta_{\rm CSC} > 50$ MeV.

In summary, in the presence of magnetic field and color superconductivity, it is the magnetic field which has the dominant effect on the structural properties of the star. Magnetic field enables the star to support more mass before it undergoes a PT to a quark core, even when the PT leads to unstable branches. In fact, the presence of magnetic field can help compensate for the softening effects of the onset of quark matter. Once a CSC quark core is formed, it is stabilized by the presence of the additional pairing energy and anisotropy of the quark phase. However, the pairing energy also serves to shift the point of PT to lower densities, resulting in more quark matter being present throughout the star, and overall softening/lower masses attained in the star. For low values of the CSC gap, i.e. $\Delta_{\rm CSC} \lesssim 50$ MeV, it is this color superconductivity softening which dominates the structure and, hence, increasing gap leads to lower $M_{\rm {\max}}$. However, for higher gaps, the effect of the color superconductivity softening can be compensated for by the presence of pressure anisotropy, particularly for Profile 2 type behavior. In highly magnetized systems, the combination of magnetic fields and CSC anisotropy can enable stars' larger CSC gaps overcome the softening effects of the large degree of quark matter and reach ``mass gap'' range masses. Hence, crucially, we show that the possibility of a deconfinement PT within NSs does not rule them out as possible mass gap candidates. If they possess high magnetic fields, they can still exhibit pressure anisotropy and subsequent mass enhancement enabling them to reach ``mass gap'' ranges.

\subsection{Anisotropy induced deformation and possibility of GWs}\label{Xsec8-5.2}

It is well known that triaxially deformed stars can emit CGWs. In this subsection, we examine the possibility of the pressure anisotropy in the HSs leading to non-zero quadrupole moment and subsequently CGW emissions. The GW calculations in this section are phenomenological estimates, based on the magnitude of deformation, as a full calculation of GW strains and associated physics requires multi-dimensional modeling. Nevertheless, we expect the results to hold qualitatively.

We assume that the presence of pressure anisotropy not only destroys spherical symmetry but also leads to triaxiality. This is in line with previous studies of magnetic field sourcing non-zero quadrupolar deformations in NSs/HSs \cite{strain_Bonazzola, GW:mag,zhang_deform,deform2}. The ellipticity $\epsilon$ of the star is a measure of this deformation. We calculate the $\epsilon$ resulting from the pressure anisotropy following a similar procedure to previous work \cite{GW:mag} as
\begin{equation}
    \epsilon = \pi/I_0 \int_V dr\ \delta\rho \ r^4,
\label{Xeqn8-5.1}
\end{equation}
where $\delta\rho$ is the density perturbation, which we model as arising from the pressure anisotropy in the star, i.e. $\sigma_{\rm aniso}$, such that $ \delta\rho = \sigma_{\rm aniso}/c^2 $ and $I_0 = (2/5)MR^2$ is the moment of inertia of the spherical star. Since we are doing an effective 1D treatment, we perform all calculations assuming an equatorial slice $\theta = \pi/2$ in the star. It is expected that the magnitude of $\delta\rho$ is, in principle, associated with a triaxial distortion, which could not be explicitly captured in the present work. Our approach, thus, gives us an approximate estimate of ellipticity. The true value of $\epsilon$, when integrated over all $\theta$, could be different from the calculated value, up to some $O(1)$ factor.

As mentioned previously, it is not enough to have just deformation for the production of CGWs - this deformation must also be \textit{triaxial}. We model the triaxiality in the form of an \textit{obliquity angle}, $\chi$, assuming that the magnetic and rotation axes of the stars are not aligned.

For a triaxially deformed star, rotating with angular velocity $\Omega \ (=2\pi\nu)$ with non-zero $\chi$ and $\epsilon$, at a distance $d$, the CGW strain produced is given by \cite{kalita:cgw},
\begin{equation}
    h_0 = \frac{4G}{c^4} \frac{\Omega^2\epsilon I_0}{d}.
\label{Xeqn9-5.2}
\end{equation}
Here, $G$ and $c$ are the gravitational constant and speed of light respectively.

CGW emission can happen at two frequencies, i.e. at $\nu$ and $2\nu$. We consider a small $\chi$ throughout, keeping in line with our 1D/spherical symmetry approximation. For small $\chi$, the GW emission at $\nu$ dominates \cite{strain_Bonazzola}. An important aspect in calculating the CGW strain is that the amplitude is divided between two different polarizations, i.e., $h_+$ and $h_\times$ \cite{strain_Bonazzola}. These polarizations are a function of $\chi$, $\Omega$ as well as the angle of inclination $i$ (the angle between the rotation axis and our line of sight). The maximum amplitude at a time $t$ is then calculated as $h = Fh_0$ \cite{kalita:cgw}, where

\begin{multline}
    F = \max \bigg| \bigg( \sin\chi \bigg[ \frac{1}{2} \cos i \sin i \cos \chi \cos\Omega t \\
    - \frac{1+\cos^2i}{2} \sin\chi \cos 2\Omega t \bigg] \bigg)\bigg|.
\label{FGW}
\end{multline}

For a given $\chi$, we can calculate the factor $F$, and the corresponding value of $i = i_{\rm max}$ that maximises it. For instance, at $t=0$, for $\chi= 5 \degree: \  F = 0.016, \ i_{\rm max} = 47.5\degree$.

Thus, given $\epsilon$ and $\chi$, one can calculate the maximum CGW strain expected to arrive at the GW detector as $h = Fh_0$. An interesting question now is: Does the additional anisotropy sourced from color superconductivity through Profiles 1 and 2 enable us to have enhanced/additional CGW signals? In other words, can CGW help us to observe the extra deformation/ellipticity that is expected from HSs with CSC quark cores, as per our model?

We restrict our study to stars with low frequency of rotation ($\nu < 30$ Hz), such that the rotation does not affect the structure of the star, in line with our 1D treatment. We study low magnetized sources, corresponding to a maximum central field of $B_0 = 10^{15} \ G$. This corresponds to a maximum surface field of $B_s = 10^{12} \ G$, following the three-order-of-magnitude scaling from previous work \cite{zuraiq}. These low fields are also expected to contribute minimally to the star's deformation and/or structure. We now examine the ellipticity produced as an effect purely of the magnetic field, i.e., $\epsilon_{\rm mag}$ arising due to $\delta\rho = B^2/8\pi c^2$, compared with the cases with pressure anisotropy (Profiles 1 and 2).

The maximum ellipticities (corresponding to $M_{\rm {\max}}$ star) due to deformation from (i) purely magnetic field effects, $\epsilon_{\rm mag}$, (ii) color superconductivity enhanced magnetic stresses of Profile 1, $\epsilon_1$, and (iii) color superconductivity driven anisotropy of Profile 2, $\epsilon_2$, as  a function of $\Delta_{\rm CSC}$ are shown in Table \ref{tab:el_GW:gap}.
Realistically, the ellipticity may be distributed around this range/order of magnitude depending on the exact stellar parameters and multi-dimensional modeling. For the stars without quark core, i.e., formed with central density below the PT density, the CSC enhancement in anisotropy is obviously absent.

\begin{table}[!htbp]
\begin{tabular}{c|c|c|c}
\hline
$\Delta_{\rm CSC}$                                                                                                                   & $\epsilon_{\rm mag}$   & $\epsilon_1$          & $\epsilon_2$         \\ \hline
 30 MeV & $8.7 \times 10^{-10}$  & $2.4 \times 10^{-9}$ & $3.1 \times 10^{-5}$ \\ 
 50 MeV & $7.1 \times 10^{-10}$ & $1.8 \times 10^{-9}$ & $2.0 \times 10^{-4}$ \\ 
 80 MeV & $8.8 \times 10^{-10}$  & $2.7 \times 10^{-8}$ & $5.1 \times 10^{-2}$ \\ \hline
\end{tabular}
\caption{\justifying{Ellipticity for different values of $\Delta_{\rm CSC}$. $\epsilon_{\rm mag}$ is the ellipticity from purely magnetic deformation, $\delta\rho = B^2/8\pi c^2$, where $B_s = 10^{12} \ G$ and $B_0 = 10^{15} \ G$, following Eq. (\ref{mag_bando}). $\epsilon_1$ is the ellipticity as a result of anisotropic deformation from Profile 1. $\epsilon_2$ is the ellipticity as a result of anisotropic deformation from Profile 2. $B_{\rm eff}$~=~100~MeV/fm$^3$ and $K_v$ = 25 GeV$^{-2}$ throughout.}}
\label{tab:el_GW:gap}
\end{table}

As Table \ref{tab:el_GW:gap}
shows, the ellipticity is enhanced from purely magnetic based deformation in both Profile 1 and Profile 2. This is seen even for low values of $\Delta_{\rm CSC}$. Since in Profile 1, the CSC gap effect is coupled with the magnetic field, it gives a smaller enhancement to ellipticity in the low field regime - leading to one-two orders enhancement. On the other hand, in Profile 2, $\Delta_{\rm CSC}$ has its own independent effect and can lead to significant enhancement of ellipticity, even leading up to $\epsilon = 5 \times 10^{-2}$ in the high gap case. Thus, even in the cases where the pressure anisotropy cannot provide adequate support against gravity and enhance the $M_{\rm {\max}}$ of the star, it can still cause adequate deformation and thus ellipticity in the star. Due to our present 1D approximation, we consider $\epsilon \lesssim 10^{-3}$ for all further analyses, which is broadly consistent with spherical symmetry \cite{mayusree:gw3}.

\begin{figure}[!htpb]
	\centering
	\includegraphics[scale=0.25]{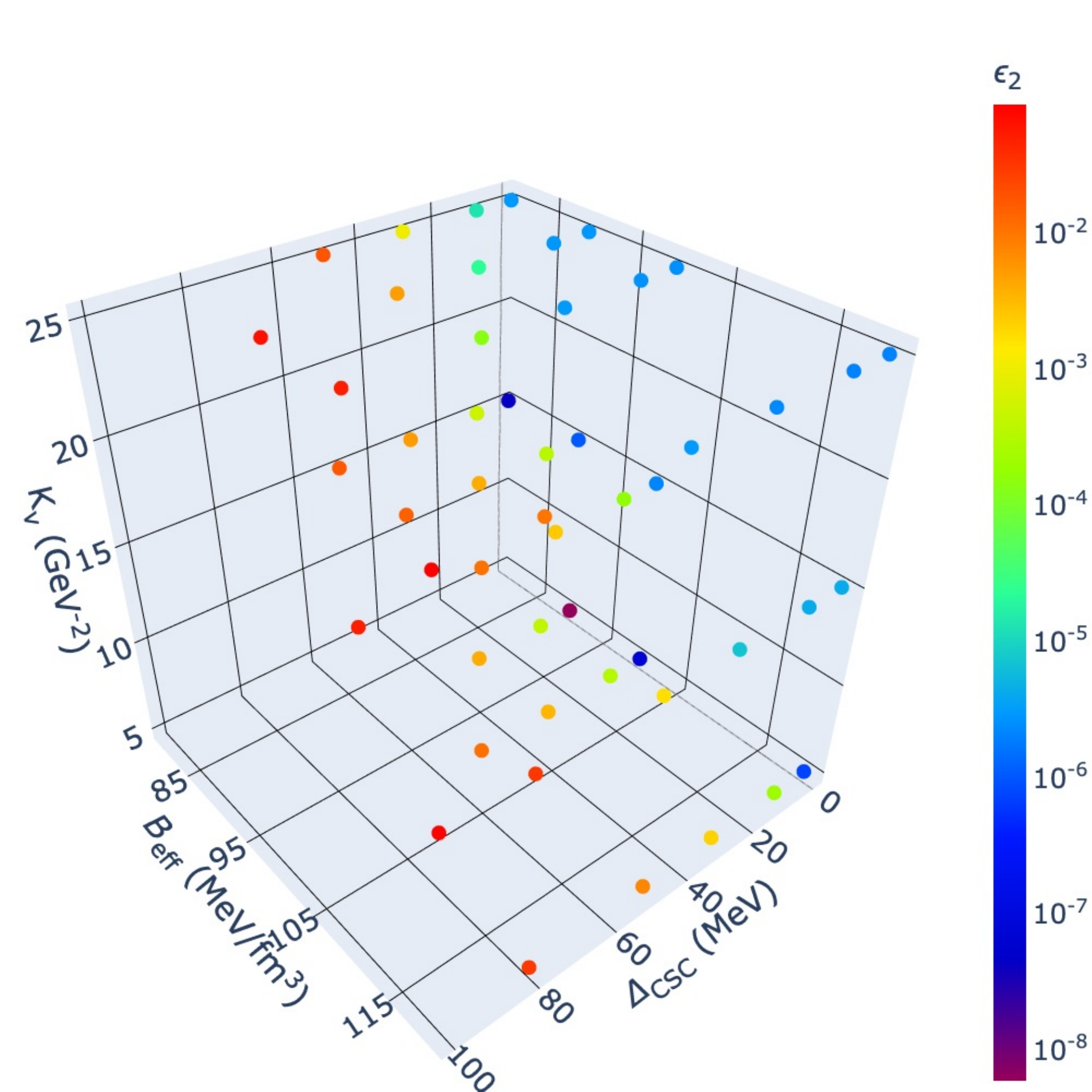}
	\caption{\justifying{Variation of ellipticity $\epsilon_2$ as a function of $\Delta_{\rm CSC}$, $B_{\rm eff}$ and $K_v$. An interactive (HTML) version is available \href{https://zeniazuraiq.github.io/ellipticity_CSCStars}{here}.}}
    \label{el_3D}
\end{figure}

$\Delta_{\rm CSC}$ is the strength of color superconductivity in the star, with an increase in $\Delta_{\rm CSC}$ indicating stronger CSC matter. This higher superconductivity is directly tied to larger deformation through the anisotropies of Profiles 1 and 2. Furthermore, all the quark matter parameters, i.e., $\Delta_{\rm CSC}$, $B_{\rm eff}$ and $K_v$, determine the size of the quark core within the HS, thus determining the extent of color superconductivity within the star, and also play a role in determining the deformation of the star leading to CGW strain. As discussed in Sec. \ref{sec:varySC}, decreasing either $B_{\rm eff}$ or $K_v$ results in a larger quark core within the star and, hence, a higher level of deformation. The variation of $\epsilon_2$ with $B_{\rm eff}$, $K_v$ and $\Delta_{\rm CSC}$ is shown in Fig. \ref{el_3D}, which is further related to CGW strain.  $\epsilon_1$ follows a similar trend but with smaller magnitudes, as the effect of CSC quark core is not prominent in this anisotropy profile.

This possible trend of enhancement of ellipticity with increasing $\Delta_{\rm CSC}$ and CSC quark core within the HS is consistent with previous work that shows how CSC quark cores can support higher strains/ellipticity \cite{ccs,gw_csc}. The resulting ellipticity being as high as  $O(10^{-2})$ is also consistent with other work \cite{ccs:gw2,ccs:gw3}. The higher level of deformation/ellipticity could possibly serve as an observational signature of CSC quark-based anisotropy, based on CGWs especially seen in Profile 2. This could also lead to a way to observationally distinguish Profile 1 and Profile 2.

Given this ellipticity, we can now estimate the resultant CGW strain and its detectability. We focus on the cases from Profile 2, where the anisotropy from color superconductivity is significant such that HSs can be deformed even in the presence of low magnetic fields. The CGW observation of a highly deformed, weakly magnetized star is thus a possible way to probe the effects of a CSC quark core inside a HS.

As mentioned previously, due to our 1D treatment, our predicted ellipticity may be slightly different when compared with a full three-dimensional (3D) calculation. Nevertheless, we calculate the strain for a range of constant ellipticities, i.e., $\epsilon = 10^{-5}$ to $\epsilon = 10^{-3}$, falling within the higher $\epsilon$ arising from the anisotropy-induced deformation of Profile 2.

We examine CGW detectability with respect to the following detectors: advanced LIGO (aLIGO), the next-generation Cosmic Explorer (CE) and Einstein Telescope (ET) detectors \cite{GW:sens}, as well as the planned Deci-Hertz Indian mission IndIGO-D \cite{mayusree_gw,mayusree_gw2}.

As mentioned previously, we need to assume some $\chi$ to perform this calculation. Realistically, $\chi$ may also affect the stellar structure, bringing with it explicit non-sphericity. Keeping these uncertainties in mind, we estimate the strain for small $\chi = 5\degree$ so that 1D based computations of moment of inertia are applicable to obtain the CGW strains predicted by our model. A full 3D non-axisymmetric model is needed to study the exact nature of deformation and to fully incorporate the effects of $\chi$ on the structure. Nevertheless, our 1D model enables us to estimate the order of magnitude of the enhanced ellipticity in Profile 2 and, thus, gives us a way to probe our model's applicability and observability. For small $
\chi$, like we consider here, the structural effects from non-sphericity can be neglected.

\begin{figure}[]
     \centering
         {\centering
         \includegraphics[width=0.5\textwidth]{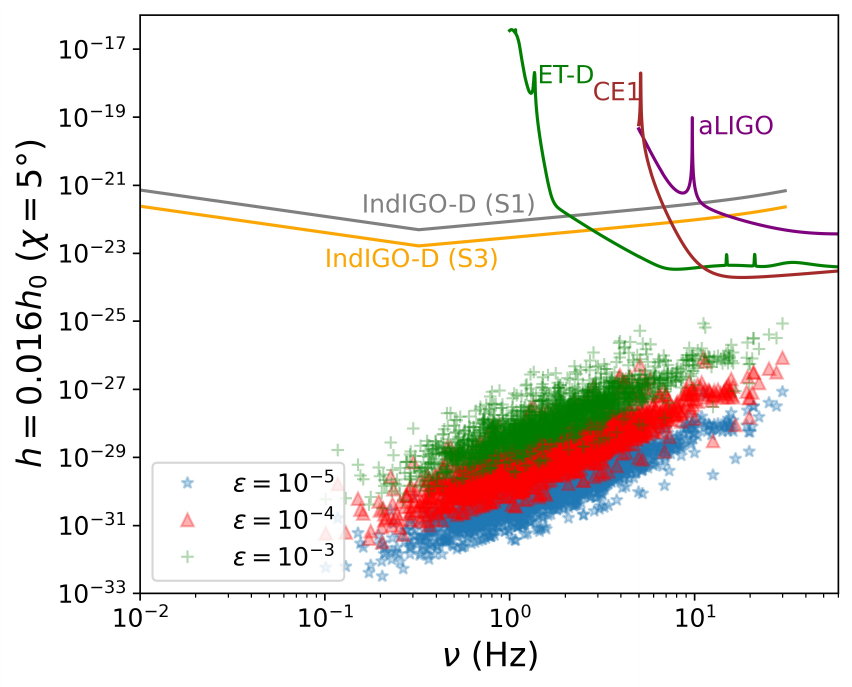}

     \vskip\baselineskip
            \includegraphics[width=0.5\textwidth]{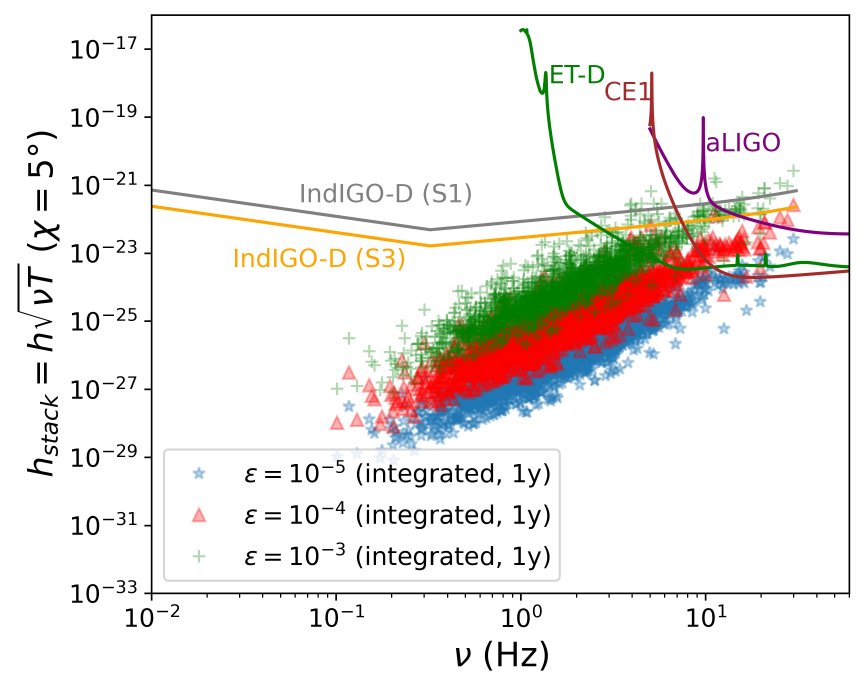}}

\caption{\justifying{The CGW strain ($h = 0.016h_0$) as a function of $\nu$ from a selection of ATNF pulsars ($\nu< 30$ Hz, $B_s \leq 10^{12}$ G) along with the characteristic strain curves for various detectors, taken from \cite{GW:sens}. Upper panel shows the instantaneous $h$.  Lower panel shows the integrated $h_{stack} = h\sqrt{\nu T}  $ after 1 year. $\chi = 5 \degree$ throughout.}}
    \label{strain_5deg}
\end{figure}

The GW strain as a function of ellipticity, calculated for different pulsars in the ATNF catalog \cite{atnf}\footnote{Available online at \url{https://www.atnf.csiro.au/research/pulsar/psrcat/}.}, is shown in Fig. \ref{strain_5deg}. We restrict our sample to pulsars with $B_s \leq 10^{12}$ G and $\nu<30$ Hz, and assume $\chi = 5\degree$, keeping in line with our previous assumptions. We further compare the strain coming from our deformed HSs to the spin-down limit\footnote{Obtained by assuming all the pulsar's spin-down luminosity is radiated through GWs. The spin-down limiting strain is calculated as \cite{ligoO2}
$$h_{\rm SD} = \frac{1}{d}\left(\frac{5GI_0}{2c^3}\frac{|\dot{\nu}|}{\nu} \right),$$ where $\dot{\nu}$ is the spin-down rate, i.e. the rate of change of the spin frequency.} \cite{ligoO2,ligoO4}. We find that for $\sim 3\%$ of our sample ATNF pulsars, the strain resulting from the highest ellipticity, i.e. $10^{-3}$, is higher than the spin-down limit, showing that the anisotropy-enhanced ellipticity may not be as high, or may be absent for those particular cases. For the cases in the ATNF catalog without a distance measurement, a spin-down limit is absent, and there is no corresponding limit on the ellipticity.

From the upper panel of Fig. \ref{strain_5deg}, we see that due to the low rotation frequencies we consider, none of the stars are observable when we consider their instantaneous strain, even when we consider $\epsilon = 10^{-3}$. However, in CGW observations, since we deal with persistent signals, it is expected that what we observe are the integrated signals. For the slow rotation periods and weak magnetic fields we consider presently, the decay of $\nu$ is not significant, and hence the CGW emission remains monochromatic. The characteristic strain for a monochromatic source is the amplitude of the wave times the square root of the number of periods observed \cite{GW:sens}, and hence, the integrated signal is $h_{stack} = h\sqrt{\nu T}$, where $T$ is the time of observation. For a few years of observation time, as is expected with the various GW missions mentioned here, we can consider the $\nu$ and $\chi$ values to be constant for a given source (see also, e.g., \cite{kalita:cgw2,mayusree:gw3}).

With 1 year of stacking, the prospects for detectability improve. We see from the lower panel of Fig. \ref{strain_5deg} that many sources could turn out to be detectable by the next-generation CE, ET and IndIGO-D detectors even with $\epsilon = 10^{-5}$.
However, none of the sources would be detected if modeled based on anisotropy Profile 1, hence they are not shown here. This is the observational difference between Profiles 1 and 2.

Our model also predicts some pulsars to be detectable using the current aLIGO detector, particularly for $\epsilon > 10^{-4}$. At the time of writing, aLIGO is on its fourth observing run O4, and has not yet seen any evidence for CGWs. aLIGO has placed CGW strain limits on around $\sim250$ pulsars across all runs \cite{ligoO2,ligoO4}, around 30 of which coincide with our sample of $\simeq 2000$ pulsars from the ATNF catalog. Considering the slowly rotating sources ($\nu< 30$ Hz) constrained by the LIGO-Virgo-KAGRA observations, the maximum ellipticity constrained from the non-detection of CGWs ranges from $10^{-3}$ to $10^{-6}$, falling within the range of ellipticities we have considered in Fig. \ref{strain_5deg}. The strictest constraint is for PSR J0453+1559 ($\nu = 21.8$ Hz, $B_s = 2.95 \times 10^9 \ G$) where the ellipticity is constrained to be $< 2.85 \times 10^{-6}$.

Thus, the non-detectability of these pulsars from the aLIGO runs up to now gives us a possible way to constrain the CSC quark parameters in our model - $B_{\rm eff}, K_v$ and $\Delta_{\rm CSC}$, especially in Profile 2. All 3 parameters together determine the size of the quark core within the HS in our model, and $\Delta_{\rm CSC}$ further determines the strength of superconductivity and, hence, ellipticity in the star, leading to a higher degree of deformation and, hence, CGW strain. For instance, the parameter combinations giving $\epsilon > 10^{-3}$ are constrained by current observations. This may possibly disfavor $\Delta_{\rm CSC} \gtrsim  50-80$ MeV, depending on the exact $B_{\rm eff}$ and $K_v$ values, as indicated in Fig. \ref{el_3D}, in the anisotropy model of  Profile 2. However, this constraint is profile dependent, and is a function of multiple parameters. More generally, this is a way to place a limit on how much CSC cores can enhance the ellipticity and/or lead to the emission of CGWs. Whether this translates to restriction of the extent of the CSC quark core itself or on its actual effect on the stellar structure requires detailed microscopic modeling. Profile 1 remains unconstrained by these possible observations, as $\epsilon_1< 10^{-8}$ in all cases. These observations are also a way to potentially distinguish between our two models, and thus more generally, between the two possible effects of superconductivity on stellar structure that we discuss in the present work.

\begin{figure}[!htb]
     \centering
         {\centering
         \includegraphics[width=0.53\textwidth]{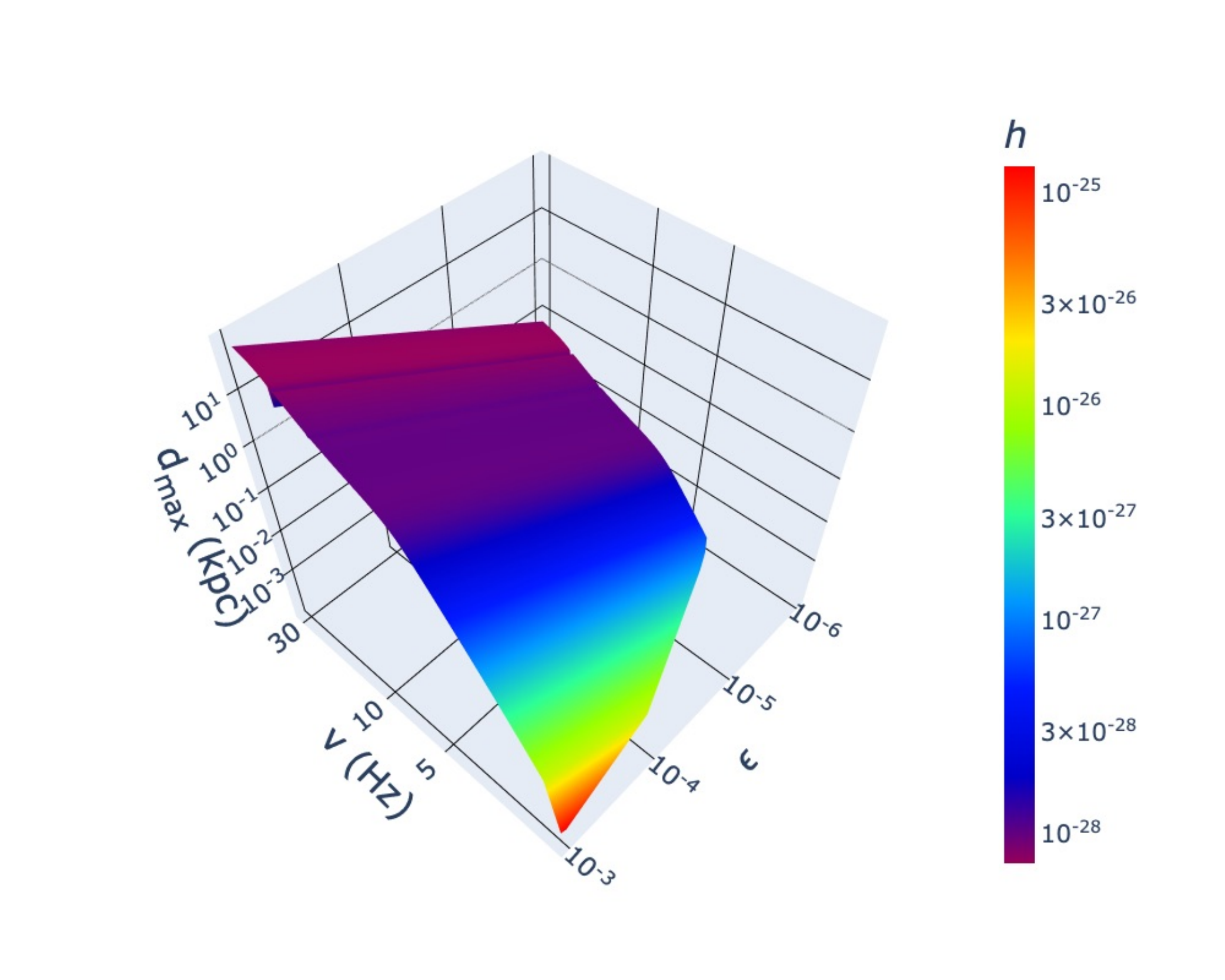}

     \vskip\baselineskip
            \includegraphics[width=0.53\textwidth]{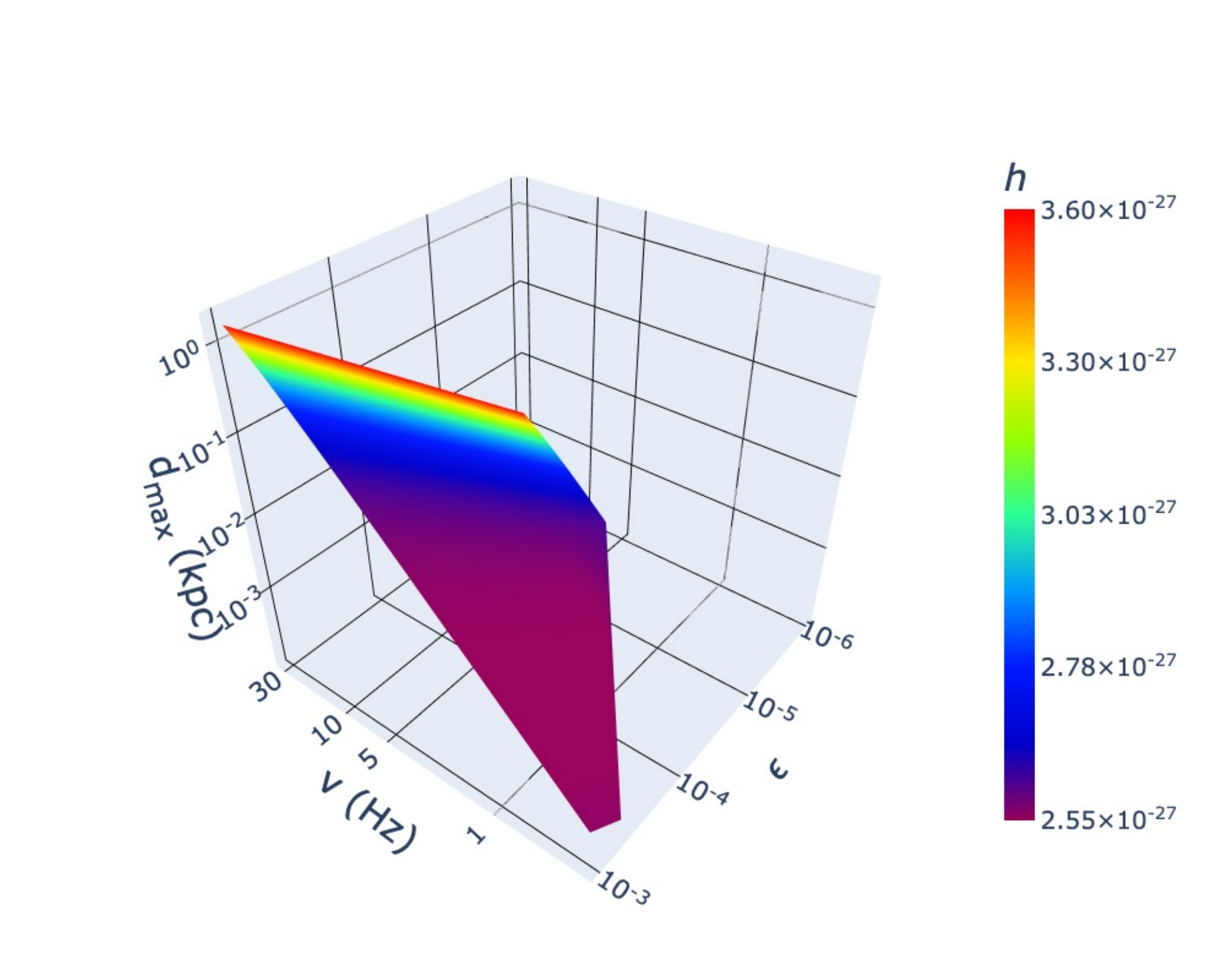}}

\caption{\justifying{$d_{\rm max}$ for ET (Upper) and IndIGO-D, S3 (Lower) detectors in a range of $\nu$ and $\epsilon$, with color indicating $h = h_{noise}/\sqrt{\nu T}$, where $T$ = 4 years. Interactive (html) versions are available at the following links: \href{https://zeniazuraiq.github.io/CGW_ET_CSCStars}{ET} and \href{https://zeniazuraiq.github.io/CGW_DeciHz_CSCStars}{IndIGO-D}.}}
    \label{maxD}
\end{figure}

Based on the above constraints, we can predict the detectability from future detectors - IndIGO-D and ET. ET is sensitive to stars with $\nu>1$ Hz, while IndIGO-D has sensitivity even below that frequency. Fig. \ref{maxD} finds out the maximum distance ($d_{\rm \max}$ ) up to which our model HS can be observed via CGW detection by these detectors as a function of $\epsilon_2$ and $\nu$, with $\chi  = 5\degree$. We assume the HS has a $M = 1.4M_\odot, R = 10$ km. We calculate $d_{\rm \max}$ as
\begin{equation}
    d_{\rm {\max}}(\nu,\epsilon) = \frac{4G}{c^4} \frac{\Omega^2\epsilon I_0}{h_{noise}}\frac{\sqrt{\nu T}}{\text{SNR}_{th}}F,
\label{Xeqn10-5.4}
\end{equation}
where $h_{noise} = \sqrt{\nu S_n(\nu)}$, $S_n(\nu)$ being the detector's power spectral density. For $\chi = 5\degree$, $F = 0.016$, as derived previously from Eq. \eqref{FGW}. $\text{SNR}_{th}$ is the threshold signal-to-noise ratio (SNR) for detection, which we take to be 12 in this case.

\begin{equation}
    d_{\rm max}(\nu,\epsilon) = \frac{4G}{c^4} \frac{\Omega^2\epsilon I_0}{h_{noise}}\frac{\sqrt{\nu T}}{\text{SNR}_{th}}F,
\end{equation}
where $h_{noise} = \sqrt{\nu S_n(\nu)}$, $S_n(\nu)$ being the detector's power spectral density. For $\chi = 5\degree$, $F = 0.016$, as derived previously from Eq. \eqref{FGW}. $\text{SNR}_{th}$ is the threshold signal-to-noise ratio (SNR) for detection, which we take to be 12 in this case.

From the $d_{\rm \max}$ shown in Fig. \ref{maxD}, we see that for slow rotating pulsars with $\epsilon = 10^{-6}$ to $10^{-3}$ arising from Profile 2, ET can detect sources upto $\simeq$ 70~kpc. This is larger than the radius of the Milky Way (= 30~kpc). Hence, deformations from a possible Profile 2-like anisotropy would be detectable over the entire sample of galactic slow-rotating pulsars (consistent with ET's science case; see \cite{ET_science}). However, IndIGO-D will be able to detect sources only upto $\simeq$ 1.5~kpc. Thus, any future detections of CGWs within the respective detector's $d_{\mathrm{max}}$ for slow-rotating and weakly magnetized sources could help us probe the ellipticity possibly arising from anisotropy model of Profile 2, and further constrain the quark parameters and/or their effect on stellar structure.

The small $\chi$ result, computed in this subsection, gives us an estimate for the \textit{lower bound} of CGW strain obtained from the possible color superconductivity enhanced ellipticity of Profile 2. Hence, the number of detectable sources may be more than what is indicated in this preliminary calculation, based on the exact distribution/value of $\chi$ in the observed pulsar population. As discussed earlier in this section, high $\chi$ cases may not be very accurately handled in a 1D approximation as non-sphericity may bring significant structural effects. However, we still expect that the changes would not be too drastic in the magnitude of $h$, as roughly estimated above, for a given $\epsilon$. Further, in this case, $\epsilon$ itself may be slightly different, due to the angular dependence of various stellar structure parameters in full 3D. In summary, the CSC enhancement of ellipticity $\epsilon_2$ by several orders in Profile 2, as opposed to the magnetically driven ellipticity models $\epsilon_{\rm mag}$ and $\epsilon_{1}$, is qualitatively established, although the exact quantitative values depend on 3D non-axisymmetric solutions and detailed microphysical modeling, not explored in the current work.

Thus, we see that if color superconductivity results in crystalline-like behavior and effects (Profile 2), it may become a source of pressure anisotropy independent of the magnetic field of the star. In the case of the slow rotating ($\nu < 30$ Hz), weakly magnetized stars ($B_s \lesssim 10^{12} \ G$) we have considered here, magnetic and rotational deformations are minimal - it is the CSC matter that leads to the potential deformation and resultant ellipticity. These are then potential CGW sources. We have made a conservative estimate of the possibility of detecting such sources by modeling with small $\chi$. Such detections (or lack thereof) prove to be an important observational constraint on CSC models in general, and can be a way to probe into the possibility of CSC quark cores inside what we observe as conventional NSs otherwise.

\section{Discussion and conclusions}\label{Xsec9-6}
\label{sec:conclusion}
The study of HSs is complex and challenging. The uncertainties of high-density matter are combined with the uncertainties associated with the hadron-quark PT. Additional physics such as color superconductivity and magnetic fields serve to make this a physically rich avenue of study. Our approach addresses the possible interplay between these different physical elements of the star, and their plausible structural and observational consequences. By combining an approximate anisotropy within a 1D framework, we are able to tackle sharp first-order PTs within HSs.

We have found that although magnetized NSs and/or magnetars may be devoid of \textit{proton} superconductivity at their cores, we can still have an emergence of color superconductivity if they contain quark cores. This may then lead to significant effects on the stellar structure and properties, as explored in the present work.

One of the challenges in studying HSs is that macroscopically, a HS containing a quark core has a very similar $M-R$ relationship/behavior as a conventional NS. This is the masquerade/camouflage effect \cite{masquerade, jaikumar_camo}. By probing pressure anisotropy, we have introduced a way to probe the internal/structural effects of color superconductivity on HS structure. This is especially seen when we turn our attention to enhanced deformation and the emission of CGWs.

We have studied the effect of (color) superconductivity at the EoS level, as well as its contributions to the pressure anisotropy of the star. We have parameterized two roles that the superconductivity can play in the generation of pressure anisotropy in the star. In Profile 1, the (proton/color) superconductivity enhances the already existing magnetic stresses in the star. This is physically motivated by type II proton superconductivity as well as CSC states such as MCFL where the superconducting gap and magnetic field are coupled together.

On the other hand, Profile 2 encapsulates the possibility where (color/proton) superconductivity leads to anisotropy independent of the magnetic field. In this case, we are motivated by anisotropy arising in superconducting/superfluid phases, such as neutron superfluidity and crystalline CSC phases.

We have found that for the mass gap range HSs, one requires high magnetic fields, and the pairing gap alone cannot bring us to mass gap ranges in either of the anisotropy profiles we have studied here. Thus, it appears that the effect of magnetic field is indispensable in order to enhance the mass of the star to ``mass gap'' levels. Notably, there is a ``color superconductivity softening'', similar to hyperon softening, that occurs due to the early onset of CSC quarks. Similar to previous work \cite{zuraiq}, we have shown that the combination of magnetic fields and anisotropy can compensate for this effect, bringing the star's $M_{\rm \max}$ to $\simeq 2.4-2.5M_\odot$. In some cases, particularly when the PT happens at a mass higher than $2M_\odot$, the magnetic field can increase the mass of hadron star attained before PT to an unstable quark branch. This leads to hadron stars with masses in the mass gap range being formed prior to the PT. Thus, the presence of a deconfinement PT does not exclude the possibility of mass gap objects. In case of stable hybrid star branches, the anisotropy can compensate for the color superconductivity softening, while in the case of unstable branches, the magnetic field can enhance the mass of the hadron star despite the PT.

Nevertheless, the presence of superconducting anisotropy, particularly from Profile 2, can bring significant ellipticity to the star, even in the weak field regimes. This is significant and brings with it implications to study the possible observational signatures, most notably, CGWs. We apply our model to the possible detectability of such CGWs, particularly in the population of slow rotating pulsars (with $\nu < 30$ Hz) with low magnetic fields ($B_s \lesssim 10^{12}$ G) such that the color superconductivity driven anisotropy from Profile 2 is the main driver of the ellipticity/deformation. The detectability (or lack thereof) of such weakly magnetized stars in current/future detectors could be an important way to probe the interiors of these stars - allowing us to answer the question of whether they contain CSC quark cores. If they do, we can further probe whether that color superconductivity can be a source of anisotropy, independent of magnetic field as we have modeled here, and has been proposed independently \cite{ccs, gw_csc}. We have made an estimate of the resultant strain and ellipticity in the current work. Extending this idea to non-axisymmetric 3D models can give us a more precise signal to look for, and a robust test of the exact role played by color superconductivity in HS structure.

This work is a first attempt to parameterize phenomenological pressure anisotropy profiles within HSs and, as such, it makes several assumptions. There are a number of directions in which this treatment can be expanded on, owing to the large parameter space/uncertainties associated with high density EoS and deconfined quark matter in general. For instance, we have relied on the vBag model - one can do similar study based on NJL type models as well. We have also considered the CSC gap to be constant in the quark phase. This is in line with the many uncertainties of the CSC phase (see, e.g. \cite{jaikumar_TB_gap}) - nevertheless, it will be interesting to see how the variation of $\Delta_{\rm CSC}$ affects the results, if at all. Additionally, we have considered Maxwell construction, i.e. local charge neutrality between the quark and hadron phases, for our PT. Imposing global charge neutrality can lead to mixed phases, i.e. Gibbs constructions - although this may not affect the $M_{\rm \max}$ significantly~\cite{jaikumar_camo}. Finally, we do not comment on the possibility of Type I vs. Type II superconductivity in the CSC sector. This is again due to the uncertain microphysics of the same, as the same definitions for distinguishing Type I vs. Type II from proton superconductivity do not carry over here~\cite{magCSC1,magCSC2}.

We leave a more microphysically motivated model incorporating the above-mentioned elements to the future work. Nevertheless, it is important to note that all models, be it microphysical or phenomenological, suffer from multiple uncertainties due to the unresolved nature of high density matter and especially the quark phase.

\section*{Acknowledgments}
The authors thank Efrain Ferrer (UTRGV), Nirmal Raj (IISc), and Fridolin Weber (SDSU) for useful discussions about the quark sector and quark/hybrid star physics. We also thank the referee for useful suggestions that have helped us improve the presentation of the work. Z.~Z.~is supported by the Prime Minister's Research Fellows (PMRF) scheme, Ref. No. TF/PMRF-22-7307. B.~M. acknowledges a project funded by SERB, India, with Ref. No. CRG/2022/003460, for partial support toward this research.

\bibliographystyle{apsrev4-1}
\bibliography{refs}
\end{document}